\documentclass[a4paper,twocolumn]{IEEEtran}
\usepackage[latin9]{inputenc}
\usepackage{color}
\usepackage{booktabs}
\usepackage{units}
\usepackage{amsmath}
\usepackage{amssymb}
\usepackage{graphicx}

\makeatletter

\pdfpageheight\paperheight
\pdfpagewidth\paperwidth

\providecommand{\tabularnewline}{\\}

\usepackage{cite}

\newcommand{\argmax}{\mathop{\mathrm{argmax}}}
\newcommand{\sinc}{\mathop{\mathrm{sinc}}}

\makeatother

\begin{document}

\title{Nonlinearity Mitigation in WDM Systems: Models, Strategies, and Achievable
Rates}

\author{Marco Secondini \IEEEmembership{Senior Member, IEEE}, Erik Agrell
\IEEEmembership{Fellow, IEEE}, Enrico Forestieri \IEEEmembership{Senior Member, IEEE},
Domenico Marsella, Menelaos Ralli Camara\thanks{M. Secondini, E. Forestieri, and M. Ralli Camara are with the Institute
of Communication, Information and Perception Technologies, Scuola
Superiore Sant'Anna, Pisa, Italy, and with the National Laboratory
of Photonic Networks, CNIT, Pisa, Italy (email: marco.secondini@sssup.it;
forestieri@sssup.it; m.rallicamara@sssup.it).}\thanks{E. Agrell is with the Department of Electrical Engineering, Chalmers
University of Technology, Gothenburg, Sweden (email: agrell@chalmers.se).}\thanks{D. Marsella is with Nokia, Vimercate, Italy (email: domenico.marsella@nokia.com).}\thanks{This paper was presented in part at the European Conference on Optical
Communication (ECOC), Gothenburg, Sweden, September 17\textendash 21,
2017 \cite{secondini:ecoc17}.}}
\maketitle
\begin{abstract}
After reviewing models and mitigation strategies for interchannel
nonlinear interference (NLI), we focus on the frequency-resolved logarithmic
perturbation model to study the coherence properties of NLI. Based
on this study, we devise an NLI mitigation strategy which exploits
the synergic effect of phase and polarization noise compensation (PPN)
and subcarrier multiplexing with symbol-rate optimization. This synergy
persists even for high-order modulation alphabets and Gaussian symbols.
A particle method for the computation of the resulting achievable
information rate and spectral efficiency (SE) is presented and employed
to lower-bound the channel capacity. The dependence of the SE on the
link length, amplifier spacing, and presence or absence of inline
dispersion compensation is studied. Single-polarization and dual-polarization
scenarios with either independent or joint processing of the two polarizations
are considered. Numerical results show that, in links with ideal distributed
amplification, an SE gain of about 1~bit/s/Hz/polarization can be
obtained (or, in alternative, the system reach can be doubled at a
given SE) with respect to single-carrier systems without PPN mitigation.
The gain is lower with lumped amplification, increases with the number
of spans, decreases with the span length, and is further reduced by
in-line dispersion compensation. For instance, considering a dispersion-unmanaged
link with lumped amplification and an amplifier spacing of 60~km,
the SE after 80 spans can be be increased from 4.5 to 4.8~bit/s/Hz/polarization,
or the reach raised up to 100 spans (+25\%) for a fixed SE. 
\end{abstract}

\section{Introduction}

In optical fiber communications, the problems of channel modeling,
nonlinearity mitigation, and capacity evaluation are strongly connected
and still open. A good channel model is needed to understand nonlinear
effects and devise effective nonlinearity mitigation strategies, with
the final goal of optimizing the modulation format and the detection
strategy. In turn, such an optimization is required to maximize the
achievable information rate (AIR) over the channel and compute its
capacity.

Ideally, a good channel model, suitable for capacity analysis, should
be: i) information-theory friendly \cite{Agrell:ITW2015}, including
an appropriate discrete representation of the input and output waveforms
and providing an explicit and mathematically tractable expression
of the conditional distribution of the output symbols given the input
ones; ii) physically accurate for a wide range of (ideally, for \emph{any})
input distributions. Unfortunately, the equations governing the propagation
of light in optical fibers\textemdash namely, the nonlinear Schrödinger
equation (NLSE), the Manakov equation, and their generalizations \cite{nl-agrawal,Wang99}\textemdash meet
only the second requirement. Thus, over the years, a number of different
approximated models of the optical fiber channel have been developed
\cite{splett1993ultimate,Peddanarappagari97,Cart:JLT99,Holzlohner:JLT02,Green2002,Van:JLT0702,KPHo:JSTQE04,Kumar:JLT0605,Serena:JLT05,Sec:JLT-CRLP,Winter:JLT2009,Pog:JLT12,Mecozzi:JLT0612,Sec:PTL2012,Beygi12,Secondini:JLT2013-AIR,Dar2013:opex,Pog:JLT14,Carena:OPEX14},
often with a conflicting view about the nature of nonlinear effects.
As a matter of fact, the available models achieve different trade-offs
between the two requirements, being either more accurate but less
information-theory friendly, or the other way around \cite{agrell17ecoc},
so that the problems of channel modeling and, in turn, of capacity
evaluation remain open \cite{Roadmap2016,Agrell20140438}.

The simplest channel models are based on the assumption that nonlinear
interference (NLI) can be accurately represented by an additive white
Gaussian noise (AWGN) term \cite{splett1993ultimate,Pog:JLT12,Beygi12,Pog:JLT14,Carena:OPEX14}.
This assumption greatly simplifies the analysis and provides a good
accuracy when applied to conventional systems. The main reason for
this accuracy becomes apparent when replacing the ambiguous concept
of ``conventional'' with the more explicit one of ``designed for
the AWGN channel''. However, extending this approach to capacity
analysis, which entails considering the best possible combination
of modulation, coding, and detection, is not likely to give the desired
results. In fact, it only provides a capacity lower bound\textemdash the
so-called \emph{nonlinear Shannon limit}\textemdash which is actually
achievable by conventional systems, but possibly very loose \cite{Agrell:jlt2015,secondini_JLT2017_scope}.\footnote{We retain the name ``nonlinear Shannon limit'' for consistency with
previous literature, although it is not a Shannon limit.} No matter how accurate AWGN-like models might seem, if we aim for
tighter capacity lower bounds or better modulation and detection strategies,
we have to drop the AWGN assumption and look for any non-AWGN channel
characteristics that can be exploited to improve system performance.
Such an attempt is made in this work.

After reviewing some popular channel models and discussing their suitability
for capacity analysis, we focus on the frequency-resolved logarithmic
perturbation (FRLP) model \cite{Sec:PTL2012,Secondini:JLT2013-AIR},
which describes interchannel NLI in optical fibers as a kind of doubly
dispersive fading (see \cite{Bello:TCOM63,liu_TrIT2004_orthogonal},
and references therein) which, at a given time and frequency, appears
mainly as \emph{phase and polarization noise} (PPN). Hence, we study
the time and frequency coherence properties of this PPN\textemdash how
fast it changes with time and frequency\textemdash and, based on it,
we devise a simple NLI mitigation strategy that relies on the synergic
effect of PPN compensation and subcarrier multiplexing (SCM) with
symbol rate optimization. The PPN evolution is modeled by a first-order
hidden Markov model (HMM), corresponding to the combination of a Wiener
phase noise (PN) and an isotropic random walk over the Poincaré sphere
\cite{Czegledi:SciRep2016}. Eventually, we verify the effectiveness
of the proposed strategy in terms of achievable spectral efficiency
(SE) and reach, resorting to a particle method for the numerical computation
of the AIR \cite{Dauwels:TrIT08}. Contrary to what is predicted by
AWGN-like models, symbol rate optimization remains effective even
for high-order modulation alphabets and Gaussian symbols if combined
with PPN compensation \cite{Mar:OFC15,dar_JLT2017_nonlinear,guiomar2017effectiveness},
which allows us to derive tighter capacity lower bounds compared to
the classical nonlinear Shannon limit. The dependence of the achievable
SE on the link length, amplifier spacing, and presence or absence
of inline dispersion compensation is also studied.

The paper is organized as follows. The system is described in Section~II.
Channel models and nonlinearity mitigation strategies are discussed
in Section III and IV, respectively. The AIR computation method is
detailed in Section V. Numerical results are presented in Section
VI. Conclusions are finally drawn in Section VII.

\section{System description\label{sec:System-description}}

\begin{figure}
\begin{centering}
\includegraphics[width=1\columnwidth]{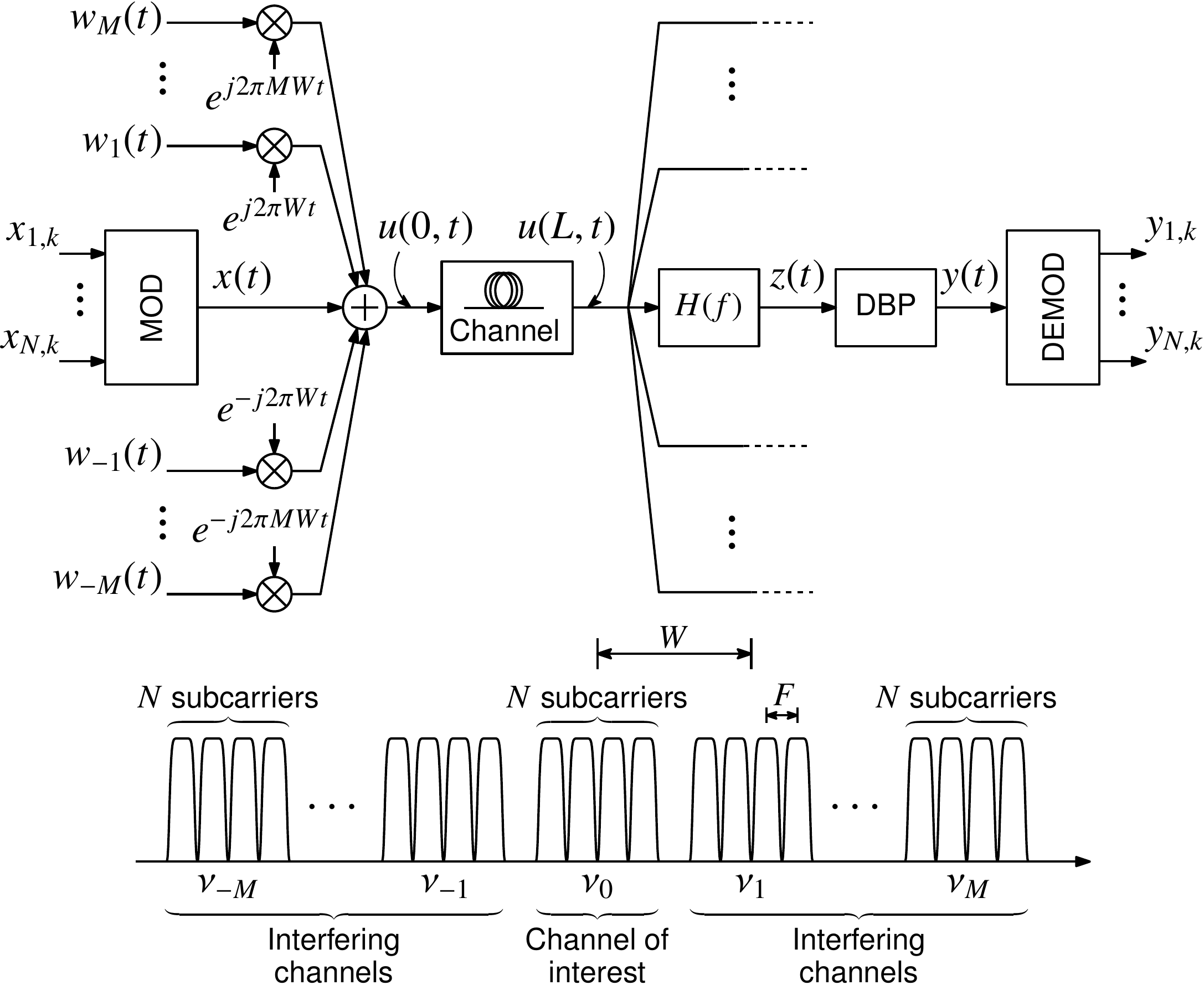}
\par\end{centering}
\caption{\label{fig:lowpass-system}Equivalent lowpass system model and corresponding
WDM spectrum.}
\end{figure}

We consider a WDM system in which $2M+1$ signals (channels)\textemdash with
channel spacing $W$ and optical center frequencies $\nu_{m}=\nu_{0}+mW$
for $m=-M,\ldots,M$\textemdash are independently generated and detected.
Each WDM channel is further decomposed into $N$ subchannels (subcarriers),
all electronically generated by the same transmitter and jointly processed
by the same receiver. The central WDM channel is arbitrarily selected
to be the channel of interest (COI), whose performance is under investigation,
while the others are considered as interfering channels (IC) affecting
the performance of the COI. The low-pass equivalent model with respect
to the COI is depicted in Fig.\,\ref{fig:lowpass-system}, where
only the block diagram for the COI has been fully represented. Here,
the $2M$ complex baseband IC signals are denoted by $w_{m}(t)$,
$m=\pm1,\ldots,\pm M$, while the COI is simply denoted by $x(t)$.
The transmitted signals can be also polarization multiplexed, in which
case they should be denoted by bidimensional vectors whose components
represent the signals on two orthogonal polarizations. For example,
the COI signal should be interpreted as\footnote{In the following, $(\cdot)^{T}$ denotes transpose, $(\cdot)^{H}$
transpose conjugate, and $(\cdot)^{*}$complex conjugate. Boldface
letters are used to denote row vectors collecting a sequence of (scalar
or dual-polarization) symbols, with a subscript denoting the number
of their elements, according to the notation $\mathbf{x}_{k}\triangleq(x_{1},\ldots,x_{k})$.
Capital sans serif letters are used to denote matrices.} $x(t)=[x_{1}(t),x_{2}(t)]^{T}$, where $x_{1}(t)$ and $x_{2}(t)$
are the aforementioned signal components on two orthogonal polarizations.
In the following, for simplicity and in order to avoid using too many
indices, all signals are to be interpreted as scalar values in the
single-polarization (1-pol) case or as bidimensional vectors in the
polarization-multiplexed (2-pol) case. When not clear from the context,
the dimensionality of the signals will be explicitly stated. Using
this convention and assuming a linear modulation format, the signal
transmitted on the COI can be written as
\begin{equation}
x(t)=\sum_{n=1}^{N}\sum_{k=1}^{K}x_{n,k}p(t-kT)e^{j2\pi f_{n}t}\label{eq:COI}
\end{equation}
where $x_{n,k}$ is the $k$-th symbol (possibly a bidimensional vector)
transmitted in the $n$-th subcarrier, $p(t)$ is the (scalar) supporting
pulse (assumed to be the same for all subcarriers and polarizations
of the COI), and $f_{n}=[n-(N+1)/2]F$, $F\le W/N$ being the frequency
separation between the subcarriers.

The COI signal (\ref{eq:COI}), which will be translated to an optical
frequency $\nu_{0}$, is then multiplexed with the IC signals generated
by the other WDM transmitters to obtain the input waveform $u(0,t)$
(see Fig.\,\ref{fig:lowpass-system}). During propagation, the WDM
signal undergoes attenuation due to fiber loss and amplification due
to optical amplifiers. So, the signal power varies along the link
according to $P(z)=a(z)P(0)$, where $P(0)$ is the launch power,
and $a(z)=P(z)/P(0)$ is the link power profile. For example, for
a link with ideal distributed amplification (IDA), $a(z)=1$; for
lumped amplification (LA), $N_{s}$ spans of the same length $L_{s}$,
and attenuation exactly compensated after each span, $a(z)=\exp[-\alpha(z\mod L_{s})]$.
In the following we will normalize the complex envelope of the propagating
signal according to the power profile. Specifically, denoting by $v(z,t)$
the complex envelope, the normalized one is $u(z,t)=v(z,t)/\sqrt{a(z)}$.
Unless otherwise stated, we assume that all WDM channels are transmitted
with the same power and use the same modulation format. Thus, the
corresponding processes are all independent and statistically equivalent
to (\ref{eq:COI}), but translated to different central frequencies
$\nu_{m}$.

After propagation through the optical fiber link, the COI is selected
by a demultiplexer filter $H(f)$, whose output $z(t)$ is processed
by digital backpropagation (DBP) (see Section~\ref{subsec:Digital-backpropagation}
for details). The backpropagated signal $y(t)$ is then sent to a
bank of $N$ matched filters to obtain the output samples
\begin{equation}
y_{n,k}=\left.y(t)\otimes\left[p^{*}(-t)e^{-j2\pi f_{n}t}\right]\right|_{t=kT}\label{eq:bank_of_matched_filters}
\end{equation}
where $\otimes$ denotes convolution. The received samples are finally
processed to recover the transmitted information according to one
of the detection strategies described in Section~\ref{subsec:Auxiliary-discrete-time-channel}.

The propagation of the normalized complex envelope $u(z,t)$ of the
WDM signal through the fiber link is governed by the NLSE in the 1-pol
case \cite{nl-agrawal}, or the Manakov equation in the 2-pol case
\cite{Wang99}. By using our convention on the signals, both can be
written as
\begin{equation}
\frac{\partial u}{\partial z}=j\frac{\beta_{2}}{2}\frac{\partial^{2}u}{\partial t^{2}}-ja(z)\gamma\Vert u\Vert^{2}u+n(z,t)\label{eq:NLSE-Manakov}
\end{equation}
where $\beta_{2}$ is the group-velocity-dispersion parameter, $\gamma$
is the nonlinear coefficient, $a(z)$ is the power profile along the
link due to attenuation or amplification, and $n(z,t)$ is the (normalized)
noise injected by optical amplifiers. The nonlinear coefficient is
expressed as $\gamma=(8/9)2\pi\nu_{0}n_{2}/(cA_{\mathrm{eff}})$,
where $n_{2}$ is the Kerr coefficient, $c$ the speed of light, $A_{\mathrm{eff}}$
the fiber effective area, and the factor of 8/9 accounts for rapid
mixing of the polarization state on the Poincaré sphere \cite{Wai:OL91}.
In the 1-pol case, $u(z,t)$ is a scalar signal and $\Vert u\Vert^{2}=|u|^{2}$.
The system parameters considered in this work are reported in Table~\ref{tab:System-parameters}.
\begin{table}
\centering{}\caption{\label{tab:System-parameters}System parameters.}
\begin{tabular}{ccc}
\toprule 
Parameter & Symbol & Value\tabularnewline
\midrule
Channel spacing & $W$ & 50~GHz\tabularnewline
Number of subcarriers & $N$ & variable\tabularnewline
Symbol time & $T$ & $T=N/W$\tabularnewline
Number of WDM channels & $2M+1$ & 5\tabularnewline
Pulse shape & $p(t)$ & $\sinc(t/T)$\tabularnewline
Attenuation & $\alpha$ & $\unit[0.2]{dB/km}$\tabularnewline
Dispersion & $\beta_{2}$ & $\unit[-21.7]{ps^{2}/km}$\tabularnewline
Nonlinear coefficient & $\gamma$ & $\unit[1.27]{W^{-1}\cdot km^{-1}}$\tabularnewline
Spontaneous emission coeff. & $\eta$ & 1 (IDA) or 1.6 (LA)\tabularnewline
\bottomrule
\end{tabular}
\end{table}

\section{Channel models\label{sec:Channel-models}}

In the following, we discuss some approximated channel models, highlighting
differences, similarities, and suitability for capacity analysis.
Moreover, we provide a more detailed description of the FRLP model,
which inspires the detection strategies described later in Section~IV.

\subsection{Split-step Fourier method (SSFM)\label{subsec:Split-step-Fourier-method}}

Rather than a channel model, the SSFM is an efficient numerical method
to solve (\ref{eq:NLSE-Manakov}) for any specific realization of
the input signal and amplifier noise. The accuracy of the SSFM can
be arbitrarily increased by reducing the step-size (for the integration
along $z$) and sampling time (for the representation of the propagating
waveform). For a given accuracy, its complexity is significantly lower
than that of other known numerical methods \cite{Taha84}. These characteristics
make the SSFM the first choice for the numerical simulation of optical
fiber systems. On the other hand, its use for information theoretical
analyses appears much more challenging, if not for the mere collection
of signal statistics. A relevant exception is the derivation of a
capacity upper bound for the optical fiber channel in \cite{Kramer2015}.
The SSFM is also used for the implementation of DBP, as discussed
in Section~\ref{subsec:Digital-backpropagation}. In this work, the
SSFM is used both for the simulation of fiber propagation according
to (\ref{eq:NLSE-Manakov}) and for the implementation of DBP.

\subsection{Regular perturbation (RP)}

Linearization techniques are often employed to tackle the study of
nonlinear effects. A widely used technique is the so-called RP method.
It consists in expanding the complex envelope $u(z,t)$ in power series
in $\gamma$, so that $u(z,t)=\sum_{k=0}^{\infty}\gamma^{k}u_{k}(z,t)$,
where $u_{0}(z,t)$ is the linear solution. Replacing this expression
in the NLSE/Manakov equation (\ref{eq:NLSE-Manakov}) and equating
the terms with equal powers of $\gamma$ that appear on both sides,
one obtains a system of recursive linear differential equations. The
solution of these equations provides $u_{k}(z,t)$ as an integral
expression containing $u_{k-1}(z,t),\ldots,u_{0}(z,t)$, so that all
$u_{k}$'s can be recursively computed starting from the linear solution
$u_{0}$. Given that an integration must be performed at each step,
the computational complexity becomes rapidly unmanageable and often
the expansion is truncated to the leading order, giving the approximation
$u(z,t)\approx u_{0}(z,t)+\gamma u_{1}(z,t)$. The term $\gamma u_{1}(z,t)$
provides an approximate description of what is commonly referred to
as \emph{nonlinear interference (NLI) noise}\textemdash though, as
we shall see, its interpretation as noise might be, in general, misleading.
In general, this NLI can be further decomposed into several different
terms, depending on the signal and/or noise components involved in
its generation. In this work, we are primarily interested in interchannel
NLI, generated by the interaction of the COI and IC signals.

Due to its simplicity, the RP method has been used in many papers
to model fiber nonlinearity \cite{Karlsson:JOSAB95,Carena:PTL97,Holzlohner:JLT02,Kumar:JLT0605,Serena:JLT05,Sec:JLT-CRLP,Mecozzi:JLT0612,Dar2013:opex}.
Moreover, it has been shown that the order $n$ RP solution coincides
with the order $2n+1$ Volterra series solution \cite{Van:JLT0702}.
Although taking into account only the leading term limits the applicability
to the realm of small nonlinearity, the RP method is very general
and, in the case of interchannel NLI, provides a simple deterministic
relationship between the first-order term $\gamma u_{1}(z,t)$ and
the IC and COI signals involved in its generation. The statistical
channel model that is obtained from this approach depends on the additional
approximations and assumptions that are commonly made to simplify
the analysis, e.g., by assuming that the perturbation term is Gaussian
or not, white or not, correlated with the COI signal or not, and so
on. For instance, starting from the RP model, the presence of a PN
component was recognized in \cite{Mecozzi:JLT0612}, a significant
time correlation of this PN component was highlighted in \cite{Dar2013:opex},
and a representation of NLI as intersymbol interference (ISI) was
given in \cite{dar2014ofc}. As we shall see, all these features are
naturally embedded in the FRLP model \cite{Sec:PTL2012,Secondini:JLT2013-AIR},
that is based on an alternative (logarithmic) perturbation method.

On the other hand, by ignoring these features and making more simplified
assumptions, the RP method leads to the GN and EGN models described
in the next subsection.

\subsection{Gaussian noise (GN) and enhanced GN (EGN) models}

A quite common trend is that of representing complicated effects,
such as fiber nonlinearity, as an additional source of AWGN. This
is the case, for instance, of the GN and EGN models \cite{Pog:JLT12,Pog:JLT14,Carena:OPEX14}.
Both models can be derived by using the RP method, with some additional
assumptions and approximations. The GN model, in particular, relies
on two main assumptions: i) NLI can be modeled as an additive Gaussian
noise, nearly white over the COI bandwidth, and independent of the
COI signal; ii) the NLI variance can be computed by assuming that,
during propagation, dispersion turns all the signals involved in the
NLI generation into Gaussian processes. These two assumptions ensure
that, given the power spectral density of the WDM signals, the optical
link can be considered as an AWGN channel, for which efficient coding,
modulation, and detection techniques are available, and system performance
and channel capacity can be computed in closed form. This approach,
besides simplifying the analysis, usually provides a good accuracy
when used to predict the performance of conventional systems.

Indeed, the second assumption entails that NLI is independent of the
modulation format of the WDM signals, while a certain dependence on
it has been observed and theoretically predicted by other models \cite{Mecozzi:JLT0612,Secondini:JLT2013-AIR}.
This is because dispersion ensures that the signal samples become
only marginally but not jointly Gaussian \cite{secondini_JLT2017_scope}.
As a result, the GN model turns out to be slightly pessimistic for
low-order modulation alphabets and, in particular, for constant-envelop
modulations. On the other hand, the EGN model drops this second assumption
and relies only on the first one. Thus, it is more complex but also
more accurate, correctly predicting the NLI dependence on modulation
format \cite{Carena:OPEX14} and allowing some nonlinear optimization
(e.g., of the symbol rate \cite{poggiolini:JLT16}).

Despite their good features, the GN and EGN models have some limitations,
so that the research of alternative models for the nonlinear optical
fiber channel is still in progress. The key factor behind these limitations
(but also behind the simplicity of the models) is the aforementioned
AWGN assumption for the NLI. In fact, making this assumption entails
that we accept a priori the impossibility of mitigating NLI and improving
the performance beyond that of conventional detectors (optimized for
the AWGN channel). This implies that the search for improved detectors,
performance, and capacity bounds (or of a proof of impossibility thereof,
should that be the case) cannot take place within the framework of
this hypothesis. And that capacity calculations, based on it, should
be interpreted only as capacity lower bounds, achievable by a mismatched
decoder optimized for the AWGN channel \cite{secondini_JLT2017_scope}.

Deviations from the AWGN model, while hard to recognize when considering
only marginal channel statistics, become more evident when considering
joint statistics \cite{secondini_JLT2017_scope}. The presence of
a PN component with a long time correlation, the dependence of NLI
on the symbols transmitted on the COI and IC signals, are all examples
of such deviations. Finding and modeling those deviations, no matter
how small they are, is the key factor to devise NLI mitigation strategies
and design improved detectors. A possible approach is discussed in
the next subsection.

\subsection{Frequency-resolved logarithmic perturbation (FRLP)\label{subsec:FRLP_model}}

Another approach to the solution of (\ref{eq:NLSE-Manakov}) is that
of replacing the nonlinear potential $\Vert u\Vert^{2}$ with a linearized
approximation \cite{mitra:nature,Green2002,Sec:PTL2012,Secondini:JLT2013-AIR}.
By rewriting the optical propagating field as the sum of a COI and
IC component $u(z,t)=s(z,t)+w(z,t)$, with $s(0,t)=x(t)$ and $w(0,t)=\sum_{m=\pm1,\ldots,\pm M}w_{m}(t)e^{j2\pi mWt}$,
and neglecting four-wave-mixing (FWM) terms, the evolution of the
COI term can be approximated as\footnote{The compact form of the nonlinear term is equivalent to the one provided
in \cite{Winter:JLT2009}, in which cross-phase modulation (XPM) and
cross-polarization modulation terms are explicitly defined. }
\begin{equation}
\frac{\partial s}{\partial z}=j\frac{\beta_{2}}{2}\frac{\partial^{2}s}{\partial t^{2}}-ja(z)\gamma(s^{H}s\,\mathsf{I}+w^{H}w\,\mathsf{I}+ww^{H})s+n(z,t)\label{eq:NLSE-Manakov-XPM}
\end{equation}
where $\mathsf{I}$ is the $2\times2$ identity matrix in the 2-pol
case, and the scalar unity in the 1-pol case. By further neglecting
the intrachannel nonlinearity term $s^{H}s$ (as it is compensated
for by single-channel DBP), and replacing the IC term $w(z,t)$ with
its linearly propagated approximation $w_{L}(z,t)$, (\ref{eq:NLSE-Manakov-XPM})
reduces to
\begin{equation}
\frac{\partial s}{\partial z}=j\frac{\beta_{2}}{2}\frac{\partial^{2}s}{\partial t^{2}}-ja(z)\gamma(w_{L}^{H}w_{L}\,\mathsf{I}+w_{L}w_{L}^{H})s+n(z,t)\label{eq:NLSE-Manakov-XPM_linearized}
\end{equation}
that is, a linear Schrödinger equation with a space- and time-variant
potential. In the 1-pol case, all the quantities in (\ref{eq:NLSE-Manakov-XPM})
and (\ref{eq:NLSE-Manakov-XPM_linearized}) become scalar, so that
$s^{H}s\,\mathsf{I}=|s|^{2}$, $w^{H}w\,\mathsf{I}+ww^{H}=2|w|^{2}$,
and $w_{L}^{H}w_{L}\,\mathsf{I}+w_{L}w_{L}^{H}=2|w_{L}|^{2}$. Finally,
assuming that the demultiplexer filter $H(f)$ removes the IC component
without affecting the COI, the received signal after single-channel
DBP can be expressed as 
\begin{equation}
y(t)=\int_{-\infty}^{\infty}\mathsf{H}_{w}(f,t)X(f)e^{j2\pi ft}df+n(t)\label{eq:linear_time-variant_model}
\end{equation}
where $\mathsf{H}_{w}(f,t)$ is a $2\times2$ time-varianttransfer
matrix, defined by a straightforward extension of the theory of linear
time-variant systems \cite{Bello:TCOM63} to the 2-pol case, and $n(t)$
is the AWGN obtained by integrating along $z$ the optical amplifier
noise $n(z,t)$. Given the characteristics of (\ref{eq:NLSE-Manakov-XPM_linearized}),
$\mathsf{H}_{w}$ must be unitary.

For the scalar (NLSE) case, the transfer matrix (which becomes a scalar
function) can be obtained from the FRLP model and expressed as \cite{Secondini:JLT2013-AIR}
\begin{equation}
\mathsf{H}_{w}(f,t)=e^{-j\theta(f,t)}\label{eq:FRLP}
\end{equation}
where the XPM term $\theta(f,t)$ depends on the IC signal $w(t)$
and link characteristics. Some general expressions for $\theta(f,t)$
and its statistics are provided in \cite{Sec:PTL2012,Secondini:JLT2013-AIR}.
Usually, the imaginary part of $\theta(f,t)$ is negligible compared
to the real one \cite{Secondini:JLT2013-AIR}, such that $\mathsf{H}_{w}(f,t)$
is unitary as required by (\ref{eq:NLSE-Manakov-XPM_linearized}).
The derivation of an analytic expression for $\mathsf{H}_{w}(f,t)$
in the 2-pol case is not considered here and will be investigated
in a future work.

\begin{figure}
\centering{}\includegraphics[width=0.75\columnwidth]{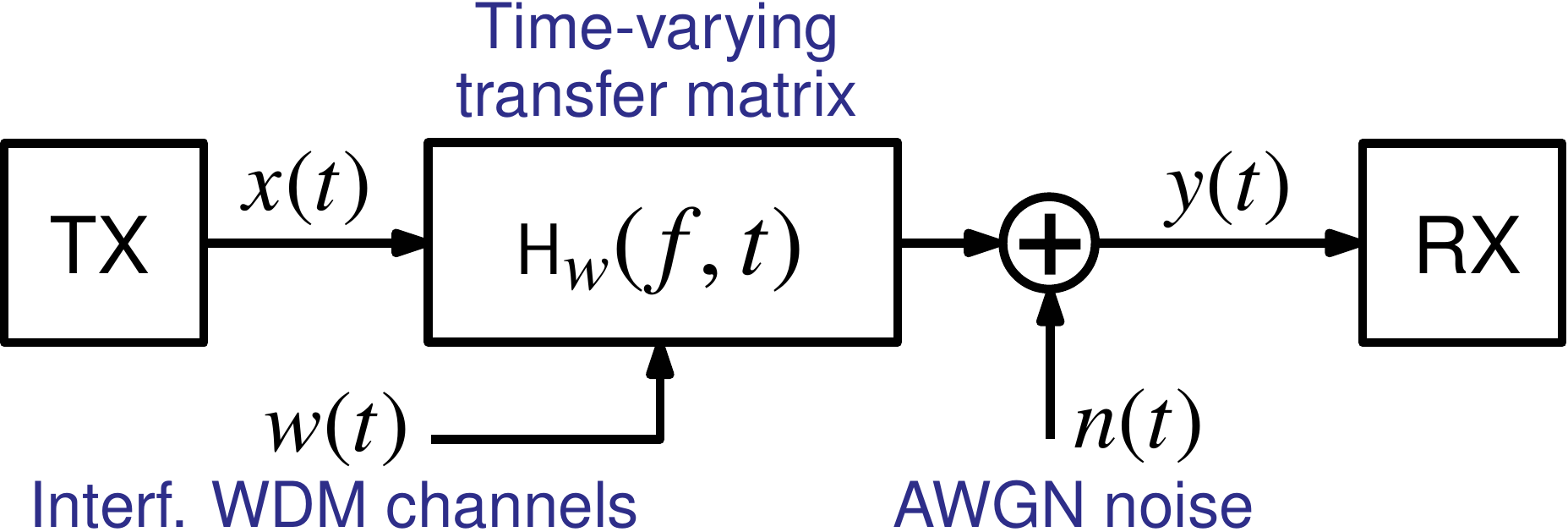}\caption{\label{fig:linearized-model}Continuous-time FRLP model. }
\end{figure}

A schematic representation of the linearized continuous-time model
in (\ref{eq:linear_time-variant_model}) is shown in Fig.~\ref{fig:linearized-model}.
 The time-variant transfer matrix $\mathsf{H}_{w}(f,t)$ depends on
the other WDM channel(s) $w(t)$ and accounts for XPM and cross-polarization
modulation. The AWGN $n(t)$ accounts for the optical amplifier noise,
whose statistics, in the linearized model (\ref{eq:NLSE-Manakov-XPM_linearized}),
remain unchanged during propagation. Other non-explicitly modeled
propagation effects are assumed to have been either compensated for
(e.g., chromatic dispersion and deterministic intrachannel NLI) or
included in the AWGN term by increasing its power spectral density
(e.g., FWM). From the COI viewpoint, the nonlinear time-invariant
optical fiber channel is seen as a linear time-variant one. This seeming
paradox is explained by the fact that channel nonlinearity is accounted
for by the dependence of $\mathsf{H}_{w}(f,t)$ on $w(t)$ \cite{Secondini:JLT2013-AIR}.
However, assuming that $w(t)$ is unknown to both the transmitter
and the receiver, such nonlinearity remains hidden, and the effect
of $\mathsf{H}_{w}(f,t)$ is simply perceived as a linear distortion.
Moreover, since $w(t)$ depends on time, also the transfer matrix
$\mathsf{H}_{w}(f,t)$ depends on time.\footnote{If $w(t)$ were known, then the distortion could be seen as nonlinear,
time-invariant, and fully deterministic, and could be exactly compensated
for, e.g., by multiplying the received signal by the inverse matrix
$\mathsf{H}_{w}^{-1}(f,t)$. This would be practically equivalent
to performing an approximated multi-channel DBP. In principle, if
$\mathsf{H}_{w}(f,t)$ were unitary\textemdash as suggested by the
approximated propagation equation (\ref{eq:NLSE-Manakov-XPM_linearized})\textemdash its
compensation would not alter the statistics of the AWGN term and would
thus completely remove the impact of nonlinearity. In this situation,
signal\textendash noise interaction, not explicitly modeled in (\ref{eq:NLSE-Manakov-XPM_linearized}),
would become dominant and limit system performance.}

The model in Fig.~\ref{fig:linearized-model} is substantially that
of a doubly dispersive fading channel, often used in wireless communications,
whose key features are the coherence time and bandwidth over which
the channel remains strongly correlated \cite{liu_TrIT2004_orthogonal}.
This analogy may help to better understand the channel characteristics
and behavior, as well as to devise improved transmission and detection
strategies. For instance, a wide coherence bandwidth (compared to
the COI) implies that the frequency dependence of $\mathsf{H}_{w}(f,t)$
can be neglected, with (\ref{eq:linear_time-variant_model}) corresponding
to a random time-variant phase and polarization rotation, here referred
to as PPN. On the other hand, a long coherence time (many symbol times)
implies that the time dependence of $\mathsf{H}_{w}(f,t)$ can be
neglected, with (\ref{eq:linear_time-variant_model}) being analogous
to a channel affected by polarization mode dispersion.

\begin{figure}
\begin{centering}
\includegraphics[width=0.514\columnwidth]{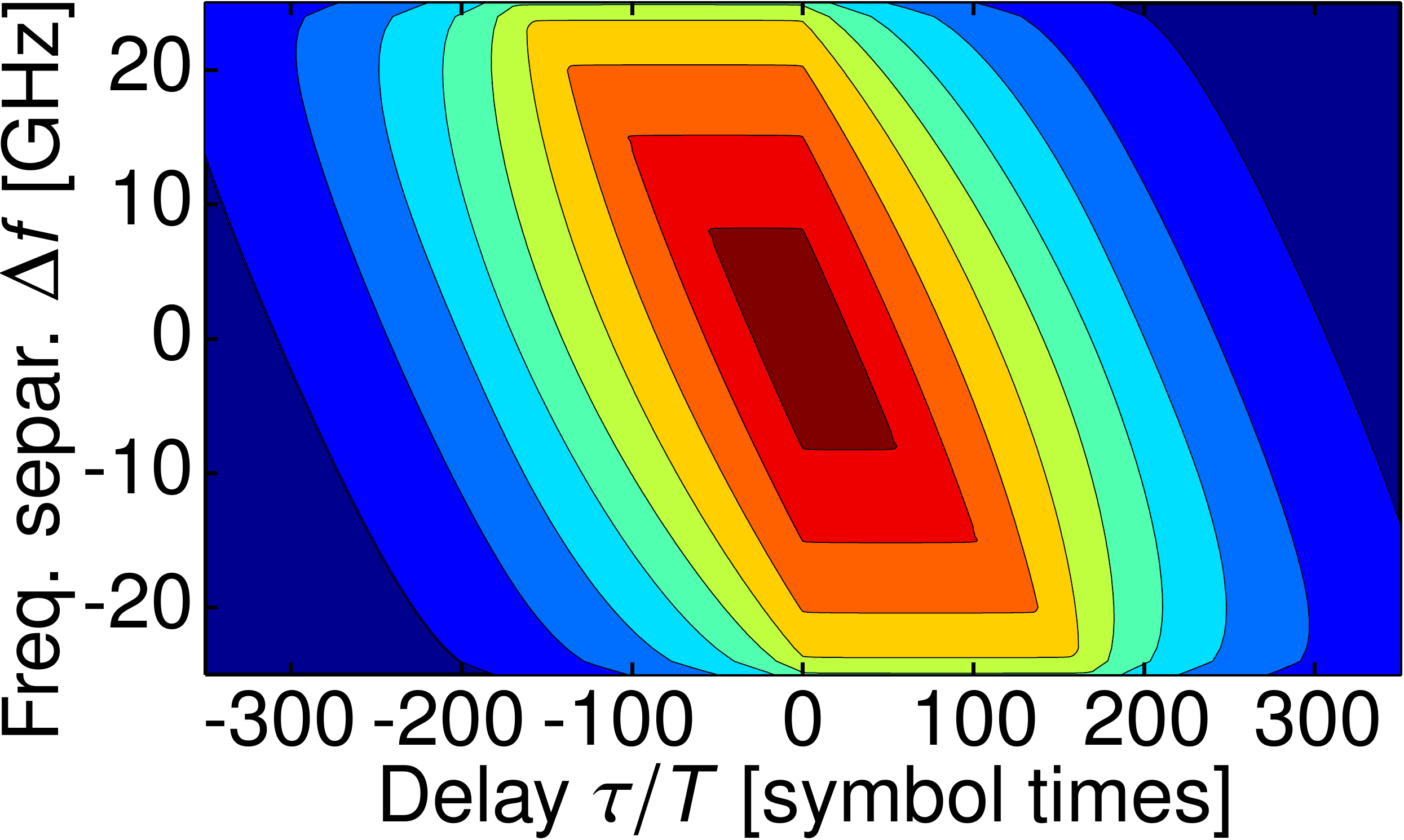}
\includegraphics[clip,width=0.47\columnwidth]{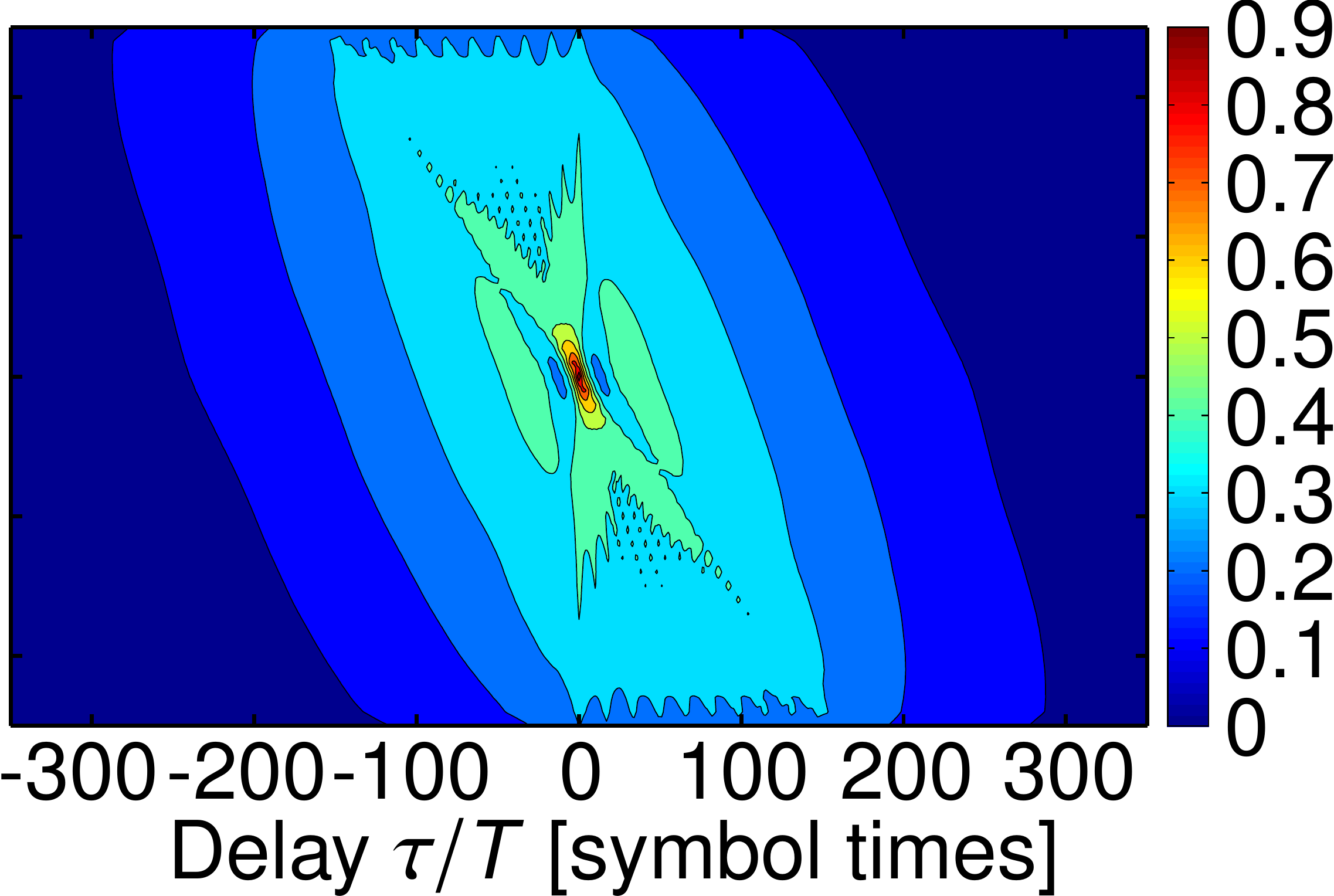}
\par\end{centering}
\hspace*{0.255\columnwidth}{\scriptsize{}(a)}\hspace*{0.43\columnwidth}{\scriptsize{}(b)}\hspace*{\fill}\caption{\label{fig:XPM-coherence}XPM coherence: (a) IDA link; (b) 10x100km
LA-DU link.}
\end{figure}

In order to illustrate the coherence properties of the channel, we
consider the 1-pol case, for which the transfer function can be analytically
approximated as in (\ref{eq:FRLP}). The contour plots in Fig.~\ref{fig:XPM-coherence}
show the correlation between the values $\theta(f,t)$ and $\theta(f+\Delta f,t+\tau)$
of the XPM term as a function of the delay $\tau$ and frequency separation
$\Delta f$.\footnote{The XPM process $\theta(f,t)$ is stationary in time but not in frequency
(its statistics depend on the frequency distance from the ICs that
generated it) \cite{Secondini:JLT2013-AIR}. Thus, the correlation
is independent of $t$ but depends on $f$, and is analyzed by holding
$f$ fixed in the middle of the COI bandwidth (conventionally set
to $f=0$), and letting $\Delta f$ vary inside it ($-W/2<\Delta f<W/2$).} Fig.~\ref{fig:XPM-coherence}(a) and (b) refer to a 1000~km IDA
link and a $10\times\unit[100]{km}$ LA dispersion unmanaged (DU)
link. Only the XPM term generated by the couple of closest ICs (i.e.,
those located at $\nu_{\pm1}=\nu_{0}\pm\unit[50]{GHz}$) is considered.
The coherence is quite substantial in the IDA link, but significantly
reduced in the LA link.
\textcolor{red}{}

\section{Nonlinearity mitigation}

In the following, we briefly discuss the nonlinearity mitigation strategies
that are relevant for the present investigation, and their relations
to the models presented in the previous section.

\subsection{Digital backpropagation\label{subsec:Digital-backpropagation}}

DBP is a channel inversion technique, obtained by a digitally-emulated
propagation of the received signal through a link with opposite coefficients
and reversed spatial configuration \cite{Essiambre:ECOC05,kahn_bp}.
According to (\ref{eq:NLSE-Manakov}), such an inversion technique
can perfectly cancel all the nonlinear effects accumulated during
propagation, with the exclusion of signal\textendash noise interaction.
Its implementation is usually based on the SSFM for its good characteristics
in terms of performance and complexity, discussed in Section~\ref{subsec:Split-step-Fourier-method}.

In principle, DBP may be either single-channel or multi-channel, depending
on how many WDM channels are jointly backpropagated. The latter has
relevant complexity issues, both in terms of hardware and processing
resources, and is implementable only when all the involved channels
are routed through the same optical path and jointly processed by
the same receiver.

In this work, we employ single-channel DBP to remove deterministic
intrachannel nonlinear effects (the term $s^{H}s$ in (\ref{eq:NLSE-Manakov-XPM}))
from the received COI signal $z(t)$. With single-channel we mean
that each WDM channel is independently backpropagated (indeed, only
the COI is actually processed in our simulations), though each WDM
channel can be further decomposed into many subcarriers that are jointly
backpropagated. The implementation of DBP is based on the SSFM. Both
the step size and sampling time are selected to obtain the best possible
performance (i.e., they are reduced until the impact of a further
reduction on the performance becomes negligible). Note that though
the expansion bandwidth over which $z(t)$ is represented equals the
inverse of the sampling time and may exceed the COI bandwidth, the
demultiplexer filter $H(f)$ ensures that only the COI is actually
backpropagated.

Some current challenges about DBP are the reduction of computational
complexity \cite{du:OE2010,secondini_PNET2016} and the inclusion
of time-varying polarization effects \cite{Czegledi:OpEx2017,liga:OFC2018}
and amplifier noise \cite{Irukulapati14tcom,irukulapati:jlt2018}.
None of those issues will be considered in this work.

\subsection{Interchannel nonlinearity mitigation}

The way in which nonlinearity is modeled has a great importance in
determining how and if its impact can be mitigated. As mentioned in
Section~\ref{sec:Channel-models}, the available models often provide
a conflicting view on the nature of nonlinear impairments. Nevertheless,
when considering a simple receiver designed for an AWGN channel, all
those models usually predict a similar performance. It does not matter
if nonlinearity generates additive noise, PN, or interference: the
impact is eventually the same and the distinction appears merely ontological.
This observation plays in favor of the simplest available models,
such as the GN and EGN ones. Yet, some models, by providing a different
view on the nature of nonlinear impairments, also suggest the possibility
of designing a better receiver that can at least partially mitigate
them. In this case, the optimal detection strategy and the corresponding
performance could actually depend on the adopted model. A trivial
but instructive example is that of intrachannel nonlinearities: if
we exclude DBP, their presence can be included in a GN-type model
and their impact on a conventional receiver correctly evaluated \cite{Pog:JLT14}.
On the other hand, by recognizing their nature of nonlinear signal
distortions induced by (\ref{eq:NLSE-Manakov}), it is possible to
compensate them almost exactly through DBP \cite{Essiambre:ECOC05,kahn_bp}.
Further improvements are also achievable by accounting for signal\textendash noise
interaction, e.g., through maximum likelihood sequence detection \cite{Mar:JLT14}
or stochastic DBP \cite{Irukulapati14tcom}.

Similar considerations hold also in the case of interchannel nonlinearities,
which are of primary interest in this work. A fairly common belief
(perhaps at the base of considering some available capacity lower
bounds as ultimate capacity limits) is that, without resorting to
joint channel processing at the transmitter or receiver, it is impossible
to mitigate interchannel NLI, which is perceived as an exogenous noise
source, outside the user control. Although it is undoubtedly true
that a complete compensation of interchannel NLI by multichannel DBP
requires joint processing, there are at least two weak points in this
kind of reasoning, which naturally lead to two different classes of
nonlinearity mitigation strategies.

The first weak point is that, as indicated by several channel models
and discussed in Section~\ref{sec:Channel-models}, interchannel
NLI is not truly equivalent to AWGN. For instance, its independence
of the COI signal is just a useful working assumption made in the
GN and EGN model to simplify the analysis. In fact, its generation
involves, to a large extent, also the COI signal, on which it depends,
causing the emergence of PPN and ISI, as discussed in Section~\ref{sec:Channel-models}.
This effects can be partly mitigated by some classical algorithms\textemdash commonly
employed also in the linear regime to counteract the ``linear version''
of this kind of impairments, e.g., due to laser PN, chromatic dispersion
and polarization mode dispersion\textemdash which can be considered
in all respects as effective nonlinearity mitigation strategies. Examples
of this first class of nonlinearity mitigation techniques, usually
operating at the receiver end, are reported in \cite{serena:ecoc2012,Sec:PTL14,dar:JLT2015,yankov:JLT2015}.
The main challenge is that channel variations induced by nonlinear
impairments are typically much faster than those observed in linear
channels, requiring algorithms with a higher adaptation speed.\footnote{As an example, typical time scales for polarization drifts due to
thermal changes and mechanical perturbations in optical fibers range
from days to milliseconds, while phase variations induced by laser
PN usually take place over microseconds (see \cite{Czegledi:SciRep2016}
and references therein). On the other hand, channel variations induced
by interchannel nonlinearity are much faster (nanoseconds), as can
be inferred from the channel coherence properties reported in Fig.~\ref{fig:XPM-coherence}.} 

The second weak point is that, though we have excluded the possibility
to deterministically control or cancel such interference by jointly
processing the COI and IC signals, it is nevertheless possible to
influence the statistics of the generated NLI by employing a proper
combination of coding and modulation. Examples of this second class
of NLI mitigation techniques, usually operating at the transmitter
end, are the design of nearly constant-envelope modulation formats
\cite{kojima:ecoc2014}; the ad hoc probabilistic shaping of QAM constellations
over one \cite{fehenberger:jlt2016} or several time slots \cite{geller:JLT2016};
the use of SCM with low-order modulation alphabets (this approach
will be further discussed in the next subsection) \cite{poggiolini:JLT16};
and the use of twin waves \cite{liu:nature2013} or conjugate data
repetition \cite{eliasson:opex2015} to partly cancel NLI. The main
challenge with these techniques is that, in most of the aforementioned
works, NLI is reduced at the expense of introducing specific shaping,
constraints, or redundancy in the transmitted signals, which often
conflict with the optimization required to maximize the SE over an
AWGN channel.

As a final remark, we note that the two classes of mitigation techniques
have been usually developed and studied separately. Therefore, it
is not guaranteed that the gains obtained by their separate use will
add up when combined in the same system. For instance, all the shaping
techniques aimed at minimizing the amplitude fluctuations of the signal
mainly remove the same NLI component that is mitigated by PN compensation,
so that little additional gains are obtained by combining the two
techniques. On the other hand, it is also possible to use the two
approaches synergistically, with an overall gain that is higher than
the sum of the gains obtained by their separate use. This possibility
will be further discussed in the next subsection.

\subsection{Subcarrier multiplexing with symbol rate optimization}

The idea of symbol-rate optimization in SCM, i.e., optimizing the
symbol rate of each subcarrier to maximize the performance, has been
investigated in several previous works \cite{behrens:ptl2011,du:opex2011,Mar:OFC15,poggiolini:JLT16,dar_JLT2017_nonlinear}.
From a theoretical point of view, and neglecting nonlinear effects,
using one or several subcarriers is perfectly equivalent in terms
of performance, given that the amplitude of the transfer function
of the fiber and of the amplifiers is almost constant over each WDM
channel bandwidth. At the same time, dividing each WDM channel into
many low-rate subcarriers may have some practical advantages and disadvantages:
it entails a natural parallelization of digital signal processing,
which runs at a lower clock rate, and provides a finer granularity
for mapping tributary channels on transport resources; but it also
requires some additional processing effort and reduces the tolerance
to laser PN. While these aspects are outside the scope of this work,
we are particularly interested in the impact of SCM and symbol rate
optimization in the presence of nonlinear effects.

One effect of decreasing the symbol rate of each subcarrier is that
of changing the relative impact of XPM and FWM, which respectively
decreases and increases, and vice versa. A possible mitigation strategy
is, therefore, that of selecting the symbol rate that provides the
best trade-off between XPM and FWM, hence minimizing the overall NLI
\cite{poggiolini:JLT16}. This effect is maximally relevant for constant-envelope
modulations, while it becomes negligible for high-order modulations
and vanishes for Gaussian modulation symbols.\footnote{For Gaussian modulation symbols, $w(0,t)$ is a Gaussian process,
whose statistical properties are fully and univocally determined by
its power spectral density. Therefore, assuming that the power spectral
density of $w(0,t)$ is independent of the number of subcarriers $N$
(this is the case with ideal rectangular Nyquist pulses), also the
statistical properties of the NLI generated by $w(0,t)$ must be independent
of $N$. In fact, the dependence of NLI on symbol rate is correctly
predicted by the EGN model, but completely absent in the GN model
\cite{poggiolini:JLT16}.} 

Changing the symbol rate has another potential effect that is not
related to the amount of generated NLI and does not vanish when considering
high-order modulation alphabets or Gaussian symbols. This effect can
be explained only by considering the non-AWGN nature of NLI, and can
be exploited only by optimizing the receiver accordingly. Indeed,
as explained in Section~\ref{subsec:FRLP_model}, the effect of NLI
can be modeled as a doubly dispersive fading, characterized by the
time-varying transfer matrix $\mathsf{H}_{w}(f,t)$ shown in Fig.~\ref{fig:linearized-model}.
The impact of $\mathsf{H}_{w}(f,t)$ on the transmitted signal and
its possible mitigation (e.g., by the compensation and equalization
strategies discussed in the previous subsection) depend on the coherence
properties of the channel\textemdash i.e., on the speed with which
$\mathsf{H}_{w}(f,t)$ changes with time and frequency\textemdash compared
to the duration and bandwidth of the supporting pulse $p(t)$ in (\ref{eq:COI}).
Such duration and bandwidth are related in a reciprocal way and scale
with the symbol rate of each subcarrier.

For instance, when a high number of low-rate subcarriers is considered,
$p(t)$ has a narrow bandwidth and a long duration. In this case,
$\mathsf{H}_{w}(f,t)$ is practically constant over each subcarrier
bandwidth, such that it can be considered as a simple phase and polarization
rotation. At the same time, it changes rapidly over time, being hard
to track and causing spectral broadening of the pulses and interchannel
interference (ICI) between subcarriers. On the other hand, when a
low number of high-rate subcarriers is considered, $p(t)$ has a large
bandwidth and a short duration. In this case, $\mathsf{H}_{w}(f,t)$
remains nearly constant over many symbol times, such that its slow
variations cause no spectral broadening and can be more easily tracked.
At the same time, it becomes highly selective in frequency, causing
a long and harmful ISI. Under such circumstances, a possible NLI mitigation
strategy is the optimization of the symbol rate to find the best trade-off
between time and frequency coherence (hence, between ISI and ICI),
making more effective the compensation of $\mathsf{H}_{w}(f,t)$ by
a detector with a limited complexity\textemdash in other words, minimizing
the mismatch between the channel and the detector.

In this work, in an attempt to maximize the AIR and find tighter capacity
lower bounds, we will only consider Gaussian modulation symbols and,
therefore, we will exploit only the latter strategy and not the XPM\textendash FWM
trade-off effect. 

\section{Achievable information rate and spectral efficiency}

The problem of computing the capacity of the discrete-time channel
in Fig.~\ref{fig:lowpass-system} is still open. The main obstacle,
as discussed in the Introduction, is the unavailability of a ``good''
channel model that explicitly provides the conditional distribution
of the output samples given the input symbols. As a consequence, the
optimum input distribution and the optimum detector (needed to achieve
channel capacity) are unknown.

The information rate of a given channel and input distribution is
the maximum bit rate at which digital information can be transmitted
using ideal error-correcting coding at an arbitrarily low error probability.
Mathematically, the information rate in bits per channel use can be
calculated using the mutual information as 
\begin{equation}
I(X;Y)\triangleq\lim_{K\rightarrow\infty}\frac{1}{K}E\Bigg\{\log\frac{p(\mathbf{y}_{K}\arrowvert\mathbf{x}_{K})}{p(\mathbf{y}_{K})}\Bigg\}\label{eq:MI}
\end{equation}
if the limit exists. Here, $X$ and $Y$ represent the discrete-time
stochastic processes of which $\mathbf{x}_{K}$ and $\mathbf{y}_{K}$,
respectively, are length-$K$ realizations. For every $K=1,2,\ldots$,
the conditional distribution $p(\mathbf{y}_{K}\arrowvert\mathbf{x}_{K})$
represents the channel law, $p(\mathbf{y}_{K})=\int p(\mathbf{y}_{K}\arrowvert\mathbf{x}_{K})p(\mathbf{x}_{K})d\mathbf{x}_{K}$
is the corresponding output distribution (connecting the input process
to the channel) and $p(\mathbf{x}_{K})$ is the selected input distribution.
The channel capacity is obtained by replacing the expectation in \eqref{eq:MI}
with the supremum of the same expectation over all input distributions
$p(\mathbf{x}_{K})$.

The capacity of realistic fiber channels, notably those represented
by the NLSE or Manakov equation \eqref{eq:NLSE-Manakov}, is not known,
neither analytically nor numerically, because no expression is known
for the corresponding channel law $p(\mathbf{y}_{K}\arrowvert\mathbf{x}_{K})$.
Therefore, optical information theory has been pursued along two main
tracks: To characterize the capacity of simplified channel models
and to establish bounds on the capacity of more realistic channel
models.

One of the most important simplified channel models is when the nonlinear
term in \eqref{eq:NLSE-Manakov} is neglected. Then the channel is
fully linear and the exact capacity is given by Shannon's famous formula
\cite{shannon48}. The linear capacity is often used as a benchmark
for more sophisticated capacity results, and it is accurate at low
launch powers. Another simplified model is obtained by neglecting
the dispersion in \eqref{eq:NLSE-Manakov}. This leads to a memoryless
channel model, which has been extensively studied under an IDA assumption.
Although the exact capacity is not known, several upper and lower
bounds have been derived \cite{Turitsyn03,yousefi11,keykhosravi17}.

With more realistic channel models, the most common approach to capacity,
and the one that will be investigated in this paper, is considering
a fixed input distribution and a mismatched detector\textemdash a
detector that is optimized for an approximated channel, known as the
\emph{auxiliary channel}, rather than for the true channel. Then the
AIR (also known as auxiliary-channel lower bound) \cite{ArLoVoKaZe06}
can be computed as\footnote{\label{fn:notation}In the following, we consider only the case in
which all the subcarriers are independently modulated and (after DBP)
detected. Thus, we focus on a generic subcarrier and, for simplicity
of notation, drop the subcarrier index.} 
\begin{equation}
I_{q}(X;Y)\triangleq\lim_{K\rightarrow\infty}\frac{1}{K}E\Bigg\{\log\frac{q(\mathbf{y}_{K}\arrowvert\mathbf{x}_{K})}{q(\mathbf{y}_{K})}\Bigg\}\label{eq:AIR}
\end{equation}
where $q(\mathbf{y}_{K}\arrowvert\mathbf{x}_{K})$ is the conditional
distribution for an arbitrary auxiliary channel and $q(\mathbf{y}_{K})=\int q(\mathbf{y}_{K}\arrowvert\mathbf{x}_{K})p(\mathbf{x}_{K})d\mathbf{x}_{K}$
is the corresponding output distribution. The expectation $E\{\cdot\}$,
on the other hand, is taken with respect to the actual (unknown) joint
distribution $p(\mathbf{x}_{K},\mathbf{y}_{K})=p(\mathbf{y}_{K}\arrowvert\mathbf{x}_{K})p(\mathbf{x}_{K})$
of the input and output processes of the true channel.

The AIR (\ref{eq:AIR}) has some important properties, which hold
for any true and auxiliary channel: i) It is a lower bound to the
information rate $I(X;Y)$ and, therefore, to channel capacity; ii)
its maximization over the auxiliary channel distribution $q(\mathbf{y}_{K}\arrowvert\mathbf{x}_{K})$
leads to the actual information rate; iii) its further maximization
over the input distribution $p(\mathbf{x}_{K})$ leads to channel
capacity; iv) it is achievable over the true channel by the maximum
a posteriori probability detector designed for the selected auxiliary
channel; and v) it can be evaluated through numerical simulations,
without an explicit knowledge of the statistics of the true channel,
provided that $q(\mathbf{y}_{K}\arrowvert\mathbf{x}_{K})$ can be
computed.

The most common auxiliary channel is the AWGN channel, in a fiber-optic
context often motivated by the GN or EGN models. Lower bounds using
this approach were presented in, e.g., \cite{Secondini:JLT2013-AIR,fehenberger15,eriksson16,liga17}.
Even some of the earliest evaluations of channel capacity \cite{splett1993ultimate,mitra:nature,wegener04}
were later identified as belonging to this class of lower bounds \cite{Agrell:ITW2015,secondini_JLT2017_scope}.
Other lower bounds were obtained by numerically estimating \eqref{eq:MI}
using uniform binary input \cite{Djordjevic05} or ring constellations
\cite{Essiambre:JLT0210}.

In contrast to the abundance of lower bounds, no upper bounds on the
capacity of \eqref{eq:NLSE-Manakov} were known until 2015, when a
closed-form upper bound on the per-sample capacity of the sampled
NLSE was presented \cite{Kramer2015}. There is however still a sizeable
gap between this upper bound and known lower bounds. This is because
the upper bound, if interpreted as bit rate or SE, is proportional
to the simulation sampling rate, which should be large for an accurate
representation of the NLSE (several samples per symbol) \cite{agrell17ecoc}.

The definitions of (achievable) information rates in (\ref{eq:MI})
and (\ref{eq:AIR}) are based on the rather abstract concept of channel
use. In practical cases, the channel is used many times by many users
(employing different time slots and frequency bands), so that a more
interesting quantity is the SE, measured in bit per second per hertz
and defined as the ratio between the information rate and the \emph{time
}and\emph{ bandwidth }required for each channel use. While required
time and bandwidth are fuzzy notions when referred to a single pulse
(channel use) \cite{Slepian:76}, in the WDM scenario of Fig.~\ref{fig:lowpass-system}
they are clearly identified by the symbol time $T$ and channel spacing
$W$ (accommodating $N$ subcarriers with spacing $F\le W/N$). In
this case, possible time and spectral broadening of the transmitted
pulses outside these values are directly accounted for by a corresponding
reduction of the information rate caused by the resulting ICI and
ISI. Thus, given the AIRs $I_{q}^{(n)}$, for $n=1,\ldots,N$, over
the $N$ different subcarriers, the achievable SE over the COI is
\begin{equation}
\mathrm{SE}_{q}=\frac{\sum_{n=1}^{N}I_{q}^{(n)}}{WT}\label{eq:spectral-efficiency}
\end{equation}
For orthogonal signaling, the minimum time-frequency separation product
between subcarriers is $FT=1$ (obtained, for instance, with rectangular
or sinc pulses). In this case, assuming no additional guard bands
between WDM channels, $WT=N$ and the achievable SE in (\ref{eq:spectral-efficiency})
(in bit per second per hertz) equals the average AIR (in bit per channel
use) over the subcarriers. This is the configuration considered in
the numerical results of Section~\ref{sec:Numerical-results}.\textcolor{magenta}{}

When computing (\ref{eq:AIR}), both the input distribution and the
auxiliary channel model can be arbitrarily selected. For what concerns
the latter, it should be selected to provide a good approximation
of the true channel, while still guaranteeing the computability of
the conditional distribution $q(\mathbf{y}_{K}|\mathbf{x}_{K})$.
Three different discrete-time auxiliary channels, based on the approximated
models described in Section~\ref{sec:Channel-models}, are considered
in the next subsection. On the other hand, the optimization of the
input distribution is not specifically addressed in this work, and
the practical choice of considering i.i.d. circularly-symmetric complex
Gaussian (CSCG) samples on each subcarrier and polarization is made.
This choice, though suboptimum at high power, is capacity-achieving
at low power (in the linear regime). Moreover, it offers some advantages
for the computation of the output distribution $q(\mathbf{y}_{K})$
(as shown in the next subsection) and represents a limit that is practically
achievable by dense quadrature amplitude modulation (QAM) constellations
with probabilistic shaping.

\subsection{Auxiliary discrete-time channel models\label{subsec:Auxiliary-discrete-time-channel}}

In the following, we consider three different discrete-time models
for the COI of Fig.~\ref{fig:lowpass-system}. The first one is a
simple AWGN model and is inspired by the GN or EGN models. The others
include some phase and/or polarization noise and are inspired by the
FRLP model. Following \cite{ArLoVoKaZe06}, we refer to these models
as \emph{auxiliary channel} models to mean that they are used to compute
the argument of the log function in (\ref{eq:AIR}) (and to design
the detector), but not compute the expectation (that is, to simulate
channel propagation), which is always done by using an accurate and
complete numerical description of the \emph{true channel} based on
the SSFM. This ensures the validity and achievability of the bound
(\ref{eq:AIR}), regardless of the assumptions and approximations\textemdash often
motivated by heuristic considerations or simply aimed at simplifying
the model\textemdash on which the auxiliary channels are based.

The auxiliary models include some constant parameters whose values
can, in principle, be derived from the equations describing the EGN
and FRLP models. Since such a derivation is not essential to our aims,
and would not guarantee the most accurate approximation of the channel,
we prefer to resort to the numerical estimation procedure described
in Section~\ref{subsec:Numerical-computation}. Such a procedure
aims at minimizing the mismatch between the true and auxiliary channel,
offering the advantage that it does not require any knowledge of the
physical channel, making it suitable also for an experimental environment,
possibly accounting for other effects which are neglected in the equations. 

All the models refer to a generic subcarrier, such that the corresponding
parameters depend on the subcarrier index. However, as explained in
footnote~\ref{fn:notation}, the subcarrier index is always omitted.
The first two models are 1-pol models, but can also be independently
applied to each polarization of a 2-pol system\textemdash in fact,
neglecting any polarization effect during propagation. The third model,
on the other hand, considers the two polarization components jointly,
and includes some polarization effects. 

\subsubsection{AWGN model\label{subsec:AWGN-model}}

When neglecting nonlinear effects, or when representing them through
the GN or EGN model, the channel of Fig.~\ref{fig:lowpass-system}
reduces to an AWGN channel. The input\textendash output relationship
for a 1-pol system and a generic subcarrier can thus be modeled as
\begin{equation}
y_{k}=cx_{k}+n_{k}\label{eq:AWGN-auxmodel}
\end{equation}
where $c=ae^{j\theta}$ is a complex constant representing a generic
attenuation/amplification $a$ and a phase rotation $\theta$, and
$n_{k}$ are i.i.d. CSCG variables with variance $\sigma_{n}^{2}$.
This is a memoryless channel, whose joint conditional distribution
factorizes into the product of the marginal distributions as
\begin{equation}
q(\mathbf{y}_{K}|\mathbf{x}_{K})=\prod_{k=1}^{K}\frac{1}{\pi\sigma_{n}^{2}}\exp\left(-\frac{|y_{k}-cx_{k}|^{2}}{\sigma_{n}^{2}}\right)\label{eq:q(y|x)_AWGN}
\end{equation}
For i.i.d. CSCG input variables with variance $\sigma_{x}^{2}$, also
the output variables are i.i.d. CSCG, with variance $\sigma_{y}^{2}=a^{2}\sigma_{x}^{2}+\sigma_{n}^{2}$,
and their joint distribution factorizes into the product of the marginal
distributions as
\begin{equation}
q(\mathbf{y}_{K})=\prod_{k=1}^{K}\frac{1}{\pi\sigma_{y}^{2}}\exp\left(-\frac{|y_{k}|^{2}}{\sigma_{y}^{2}}\right)\label{eq:output-auxdistribution}
\end{equation}

\subsubsection{PN model\label{subsec:Phase-noise-model}}

A more accurate representation of XPM can be obtained by using the
FRLP model (\ref{eq:FRLP}). The frequency dependence of the XPM term
$\theta(f,t)$ within each subcarrier bandwidth and its time dependence
within a pulse duration cause ISI and ICI, respectively. By neglecting
such ISI and ICI\textemdash whose minimization is the purpose of optimizing
the number of subcarriers\textemdash the input\textendash output relationship
for a 1-pol system and a generic subcarrier can be modeled as
\begin{equation}
y_{k}=ae^{j\theta_{k}}x_{k}+n_{k}\label{eq:phase-noise-auxmodel}
\end{equation}
where $a$ and $n_{k}$ are as in the AWGN channel (\ref{eq:AWGN-auxmodel}),
and $\theta_{k}$ is the $k$-th sample of a discrete-time process
that accounts for XPM. In principle, the statistics of the XPM process
could be derived from the FRLP equations. In particular, an important
property of the process is its long temporal coherence which, as shown
in Fig.~\ref{fig:XPM-coherence}, might span many symbol times. Moreover,
as explained in \cite{Secondini:JLT2013-AIR}, the process can be
approximately assumed to be real-valued and have Gaussian samples.
For the sake of simplicity, we model it as a random walk (the discrete-time
analog of a Wiener process) on a circle
\begin{equation}
\theta_{k}=\theta_{k-1}+\Delta\theta_{k}\,\mod\,2\pi\label{eq:random-walk-PN}
\end{equation}
 with i.i.d. real zero-mean Gaussian increments $\Delta\theta_{k}$
with variance $\sigma_{\theta}^{2}$.

The resulting channel model (\ref{eq:phase-noise-auxmodel}) is a
first-order HMM, with $\theta_{k}$ playing the role of the (hidden)
state variable and taking continuous values in the range $[0,2\pi)$.
In this case, the conditional distribution does not factorize into
the marginal distributions, but can be expressed as
\begin{equation}
q(\mathbf{y}_{K}|\mathbf{x}_{K})=\int q(\mathbf{y}_{K},\theta_{K}|\mathbf{x}_{K})d\theta_{K}\label{eq:q_conditional_final}
\end{equation}
where the term in the integral is recursively computed as
\begin{multline}
q(\mathbf{y}_{k},\theta_{k}|\mathbf{x}_{k})=\\
\int q(\mathbf{y}_{k-1},\theta_{k-1}|\mathbf{x}_{k-1})q(y_{k}|x_{k},\theta_{k})q(\theta_{k}|\theta_{k-1})d\theta_{k-1}\label{eq:q_conditional_recursion}
\end{multline}
for $k=1,\ldots,K$, with
\begin{align}
q(y_{k}|x_{k},\theta_{k}) & =\frac{1}{\pi\sigma_{n}^{2}}\exp\left(-\frac{|y_{k}-ae^{j\theta_{k}}x_{k}|^{2}}{\sigma_{n}^{2}}\right)\label{eq:q(y|x,theta)}\\
q(\theta_{k}|\theta_{k-1}) & =\frac{1}{\sigma_{\theta}\sqrt{2\pi}}\exp\left(-\frac{(\theta_{k}-\theta_{k-1})^{2}}{2\sigma_{\theta}^{2}}\right)\label{eq:theta_transition_probability}
\end{align}
No prior knowledge about $\theta_{0}$ is assumed, so the recursion
is initiated by setting $q(\mathbf{y}_{0},\theta_{0}|x_{0})\triangleq q(\theta_{0})$
uniform in $[0,2\pi)$. On the other hand, considering i.i.d. CSCG
input variables, the output distribution is not affected by the presence
of PN and can be expressed as in (\ref{eq:output-auxdistribution}).

The random walk in (\ref{eq:random-walk-PN}) is the same model commonly
adopted for laser PN. In fact, (\ref{eq:random-walk-PN}) could even
represent the combination of XPM and laser PN, with $\sigma_{\theta}^{2}$
being the sum of the variances related to the two processes. This
is, however, a clear oversimplification of the actual evolution of
the XPM term. In a random walk, the process value at a given time
is equally determined by all the previous steps, so that its standard
deviation keeps increasing with time. When the random walk is wrapped
to a finite interval\textemdash the $[0,2\pi)$ interval in our case\textemdash this
yields a uniform distribution of the process value after a few steps.
While this is a reasonable assumption for laser PN, it is not so for
XPM. In fact, the XPM term at a given time is determined only by the
values of the IC signals over a finite window around that time, so
that its standard deviation remains constant over time\textemdash in
practice, the phase fluctuates in a small interval around its mean
value. Such a behavior could be modelled by adding a mean-reverting
term (proportional to the difference between the mean and current
values of the phase) to (\ref{eq:random-walk-PN}), obtaining a first-order
autoregressive (AR) process. Eventually, a higher-order AR process
could be employed to better account for the actual XPM statistical
properties\textcolor{red}{} and for the possible combination of XPM
and laser PN. In this work, the choice of the simple (but clearly
mismatched) model in (\ref{eq:random-walk-PN}) is justified a posteriori
by the performance gains shown in Section~\ref{sec:Numerical-results}.
Possible improvements achievable by using more refined models will
be investigated in a future work.

\subsubsection{PPN model\label{subsec:Phase-and-polarization}}

The model (\ref{eq:phase-noise-auxmodel}) can be generalized to account
for polarization effects induced by the Manakov equation (\ref{eq:NLSE-Manakov}).
In this case, considering two-dimensional input and output vectors
$x_{k}$ and $y_{k}$, and assuming that the overall effect of the
channel is still that of inducing a unitary transformation and adding
AWGN as in (\ref{eq:linear_time-variant_model}), the input\textendash output
relationship for a generic subcarrier can be modeled as
\begin{equation}
y_{k}=ae^{j\theta_{k}}\mathsf{J}_{k}x_{k}+n_{k}\label{eq:PPN-auxchannel}
\end{equation}
where $a$ and $\theta_{k}$ are as in the PN model (\ref{eq:phase-noise-auxmodel}),
$n_{k}$ is a two-dimensional noise vector, whose elements are independent
and distributed as in the AWGN and PN model, and $\mathsf{J}_{k}$
is a $2\times2$ complex unitary Jones matrix that accounts for a
generic rotation of the state of polarization. In analogy to the random-walk
model adopted in (\ref{eq:random-walk-PN}), the evolution of $\mathsf{J}_{k}$
is modeled as an isotropic random walk over the Poincaré sphere \cite{Czegledi:SciRep2016}
\begin{align}
\mathsf{J}_{k} & =\exp(j\boldsymbol{\alpha}_{k}\cdot\boldsymbol{\vec{\sigma}})\mathsf{J}_{k-1}\nonumber \\
 & =\exp\left(\begin{array}{cc}
j\alpha_{1,k} & \alpha_{3,k}+j\alpha_{2,k}\\
-\alpha_{3,k}+j\alpha_{2,k} & -j\alpha_{1,k}
\end{array}\right)\mathsf{J}_{k-1}\label{eq:rotation_matrix}
\end{align}
where $\boldsymbol{\vec{\sigma}}$ is the Pauli spin vector in Stokes
space \cite{gordon:pnas2000}, and the real innovation parameters
$\boldsymbol{\alpha}_{k}=(\alpha_{1,k},\alpha_{2,k},\alpha_{3,k})$
are drawn independently from a zero-mean Gaussian distribution with
variance $\sigma_{p}^{2}$ at each time instance $k$.

The obtained model is still a first-order HMM, whose hidden state
is represented by the pair $(\theta_{k},\mathsf{J}_{k})$. The joint
conditional distribution $q(\mathbf{y}|\mathbf{x})$ can thus be expressed
as in (\ref{eq:q_conditional_final})\textendash (\ref{eq:q_conditional_recursion}),
but replacing $\theta_{k}$ with $(\theta_{k},\mathsf{J}_{k})$ and
extending integrations to the corresponding state space. The conditional
distribution (\ref{eq:q(y|x,theta)}) must be replaced by
\begin{equation}
q(y_{k}|x_{k},\theta_{k},\mathsf{J_{k}})=\frac{1}{\pi^{2}\sigma_{n}^{4}}\exp\left(-\frac{\parallel y_{k}-ae^{j\theta_{k}}\mathsf{J}_{k}x_{k}\parallel^{2}}{\sigma_{n}^{2}}\right)\label{eq:q(y|x,theta,J)}
\end{equation}
and the transition probability $q(\mathsf{J}_{k}|\mathsf{J}_{k-1})$
could be derived from (\ref{eq:rotation_matrix}).\footnote{For the particle method adopted in the next section to compute (\ref{eq:q_conditional_final})\textendash (\ref{eq:q_conditional_recursion}),
an explicit expression of the transition probability is not required,
as the simple algorithmic representation based on (\ref{eq:rotation_matrix})
is sufficient.} As in the 1-pol case, for i.i.d. CSCG input symbols, the output distribution
is not affected by the unitary transformation in (\ref{eq:rotation_matrix})
and can be expressed as the product of two distributions as in (\ref{eq:output-auxdistribution}),
one per each polarization component.

We remark that the same considerations expressed at the end of Section~\ref{subsec:Phase-noise-model}\textemdash the
mismatch between the random-walk model (\ref{eq:random-walk-PN})
and the actual PN evolution, and the better modelling options offered
by AR processes\textemdash hold also for (\ref{eq:rotation_matrix})
and the actual PPN evolution.

\subsection{Numerical computations\label{subsec:Numerical-computation}}

The problem of computing the AIR (\ref{eq:AIR}) and the corresponding
achievable SE (\ref{eq:spectral-efficiency}) cannot be solved analytically,
as the joint input\textendash output distribution $p(\mathbf{x}_{K},\mathbf{y}_{K})$
of the true channel\textemdash with respect to which the expectation
in (\ref{eq:AIR}) must be computed\textemdash is generally unknown.
However, an accurate numerical estimate can be obtained by relying
on the asymptotic equipartition property \cite[Ch.~3]{cover06} and
following the procedure described in \cite{ArLoVoKaZe06}. One additional
difficulty is that the HMMs in (\ref{eq:phase-noise-auxmodel}) and
(\ref{eq:PPN-auxchannel}) are characterized by a continuous state
space, such that the recursive computation of (\ref{eq:q_conditional_recursion})
would require the computation of a long series of nested integrals.
Among several possible approaches to address this issue\footnote{A typical approach is the quantization of the state space which, however,
becomes infeasible in high-dimensional spaces\textemdash in fact,
it provides good results for the monodimensional state space of (\ref{eq:phase-noise-auxmodel})
but is already too complex for the four-dimensional state space of
(\ref{eq:PPN-auxchannel}). An alternative low-complexity approach
is the use of properly selected parametrized canonical distributions
(e.g., Gaussian or Tikhonov) to represent (\ref{eq:q_conditional_recursion}),
but it usually entails some approximations and must be specifically
developed and tested for each model. For a general discussion, see
\cite{Colavolpe:JSAC05} and references therein.}, we resort to the particle method proposed in \cite{Dauwels:TrIT08},
because it can be easily extended to general HMMs, its accuracy can
be arbitrarily increased by increasing the number of particles, and
its complexity scales fairly well with the dimensionality of the state
space. The overall procedure for the computation of (\ref{eq:AIR})
(for a 1-pol system, a single subcarrier, and the PN model) is briefly
summarized in the following.
\begin{enumerate}
\item Draw two long input sequences $\mathbf{x}_{K}=(x_{1},\ldots,x_{K})$
and $\mathbf{x}'_{K'}=(x'_{1},\ldots,x'_{K'})$ of i.i.d. CSCG samples.
\item Compute the corresponding output sequences $\mathbf{y}_{K}=(y_{1},\ldots,y_{K})$
and $\mathbf{y}'_{K'}=(y'_{1},\ldots,y'_{K'})$ by using the SSFM
to simulate the true channel.
\item \label{enu:Estimate-the-parameters}Estimate the parameters $a$,
$\sigma_{n}^{2}$, $\sigma_{\theta}^{2}$ of the auxiliary channel
(\ref{eq:phase-noise-auxmodel}) by using the sequences $\mathbf{x}'_{K'}$
and $\mathbf{y}'_{K'}$.
\item Compute the conditional distribution $q(\mathbf{y}_{K}|\mathbf{x}_{K})$
in (\ref{eq:q_conditional_final}) by using the particle method \cite{Dauwels:TrIT08}.
In particular, for $k=1,\ldots,K$:
\begin{enumerate}
\item Generate the new particle set for the representation of $\theta_{k}$
by drawing samples from (\ref{eq:theta_transition_probability}).
\item Update the state metric for each particle by using (\ref{eq:q(y|x,theta)}).
\item Use normalization and resampling \cite{Dauwels:TrIT08} to avoid numerical
issues.
\end{enumerate}
\item Compute the corresponding output distribution $q(\mathbf{y}_{K})$
by using (\ref{eq:output-auxdistribution}).
\item Estimate the AIR as
\begin{equation}
\hat{I}_{q}(X;Y)=\frac{1}{K}\log\frac{q(\mathbf{y}_{K}\arrowvert\mathbf{x}_{K})}{q(\mathbf{y}_{K})}
\end{equation}
and the corresponding achievable SE in (\ref{eq:spectral-efficiency}).
\end{enumerate}
Some remarks follow.
\begin{itemize}
\item For the AWGN model, the particle approach is not required, as the
conditional and output distributions can be directly computed from
(\ref{eq:q(y|x)_AWGN}) and (\ref{eq:output-auxdistribution}), respectively.
\item For 2-pol systems, the AWGN or PN models can be independently applied
to each polarization component of each subcarrier. Moreover, the PPN
model can be jointly applied to vector samples, with a straightforward
extension of the described numerical procedure, i.e., including a
particle representation of the matrix $\mathsf{J}_{k}$, replacing
(\ref{eq:theta_transition_probability}) with (\ref{eq:rotation_matrix})
and (\ref{eq:q(y|x,theta)}) with (\ref{eq:q(y|x,theta,J)}), and
estimating the additional parameter $\sigma_{p}^{2}$ of the PPN model.
\item In Step~\ref{enu:Estimate-the-parameters}, channel parameters are
estimated by using a two-step procedure. Given each input variable
$x'_{k}$, the output squared modulus $\Vert y'_{k}\Vert^{2}$ has
a noncentral chi-squared distribution that does not depend on the
presence of PPN. Thus, in the first step, $a$ and $\sigma_{n}^{2}$
are estimated by numerically maximizing the corresponding likelihood
\begin{align}
\hat{\sigma_{n}^{2}} & =\argmax_{\sigma_{n}^{2}}\sum_{k=1}^{K'}\log\left[\frac{1}{\sigma_{n}^{2}}\exp\left(-\frac{\Vert y'_{k}\Vert^{2}+a^{2}\Vert x'_{k}\Vert^{2}}{\sigma_{n}^{2}}\right)\right.\nonumber \\
 & \left.\left(\frac{\Vert y'_{k}\Vert^{2}}{a^{2}\Vert x'_{k}\Vert^{2}}\right)^{p}I_{2p}\left(\frac{2a\Vert x'_{k}\Vert\Vert y'_{k}\Vert}{\sigma_{n}^{2}}\right)\right]
\end{align}
where $a=\sqrt{(\sigma_{y}^{2}-\sigma_{n}^{2})/\sigma_{x}^{2}}$;
$\sigma_{x}^{2}$ and $\sigma_{y}^{2}$ are the variances of the input
and output samples, respectively; $p=0$ or 0.5 for the 1-pol and
2-pol case, respectively; and $I_{m}(\cdot)$ is the modified Bessel
function of the first kind of order $m$. In the second step, the
remaining parameter $\sigma_{\theta}^{2}$ (and, in the 2-pol case,
$\sigma_{p}^{2}$) is estimated by numerically maximizing the AIR,
which corresponds to minimizing the mismatch between the true and
auxiliary channels.
\item For systems with $N$ subcarriers, $N$ independent input sequences
are drawn to generate the input signal (\ref{eq:COI}). The corresponding
output signal is computed by using the SSFM, and the $N$ output sequences
are then extracted by the bank of matched filters according to (\ref{eq:bank_of_matched_filters}).
The AIR is finally computed independently for each subcarrier and
averaged according to (\ref{eq:spectral-efficiency}).
\end{itemize}

\section{Numerical results\label{sec:Numerical-results}}

In this section, by following the procedure detailed in Section~\ref{subsec:Numerical-computation},
we compute the achievable SE for the COI of the system described in
Section~\ref{sec:System-description} under constrained modulation
(CSCG input symbols) and mismatched decoding, considering the system
parameters provided in Table~\ref{tab:System-parameters}, 1-pol
and 2-pol transmissions, and different link configurations. We compare
the SE that can be achieved by using three different detectors, matched
to the AWGN, PN, and PPN auxiliary channel models provided in Section~\ref{subsec:Auxiliary-discrete-time-channel},
and by optimizing the number of subcarriers (or equivalently, the
symbol rate per subcarrier). For the considered pulse shape, according
to (\ref{eq:spectral-efficiency}), the achievable SE over the COI
equals the average AIR over the subcarriers. DBP is implemented as
explained in Section~\ref{subsec:Digital-backpropagation}. SE values
and channel parameters are estimated by averaging over $K=100000$
and $K'=10000$ symbols per subcarrier, respectively. All the plots
report the SE per polarization.

\begin{figure*}
\includegraphics[width=0.25\textwidth]{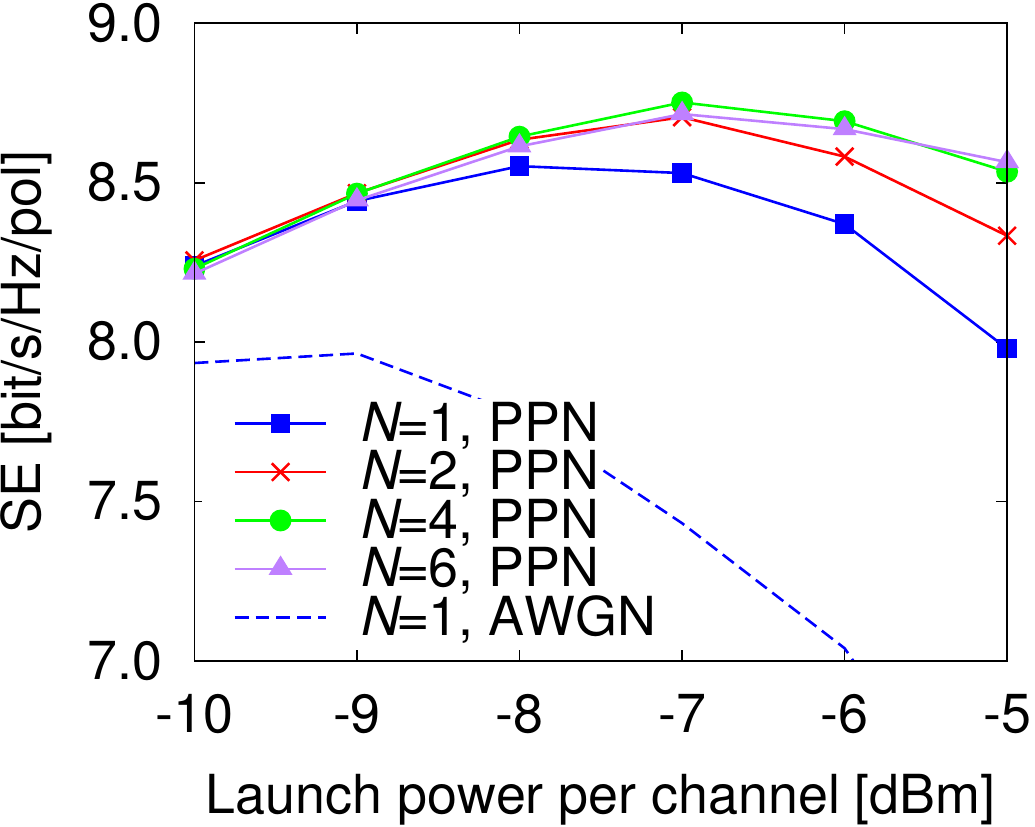}\includegraphics[width=0.25\textwidth]{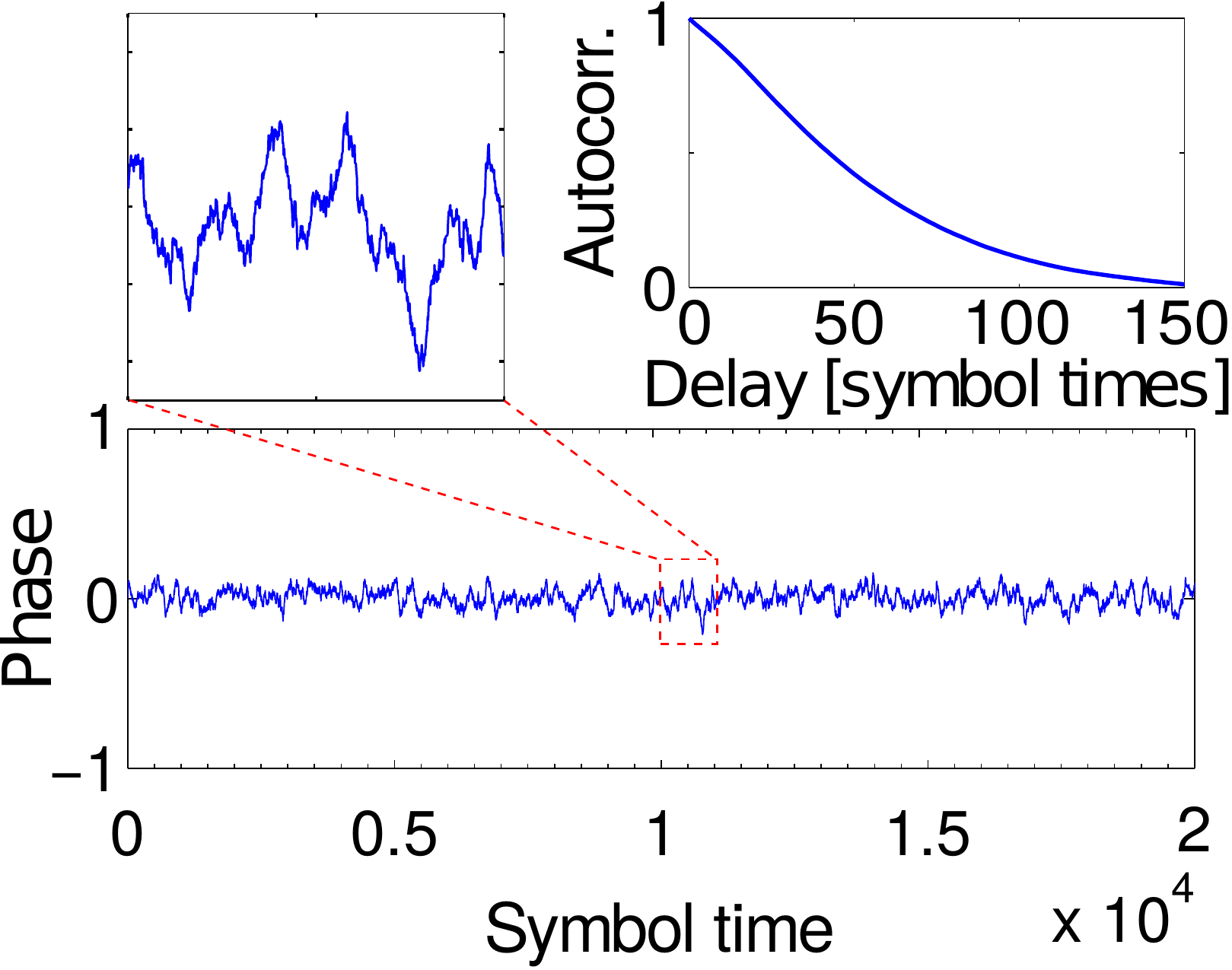}%
\begin{minipage}[b][22ex][t]{0.25\textwidth}%
\begin{center}
\includegraphics[viewport=100bp 100bp 950bp 780bp,clip,width=0.8\columnwidth]{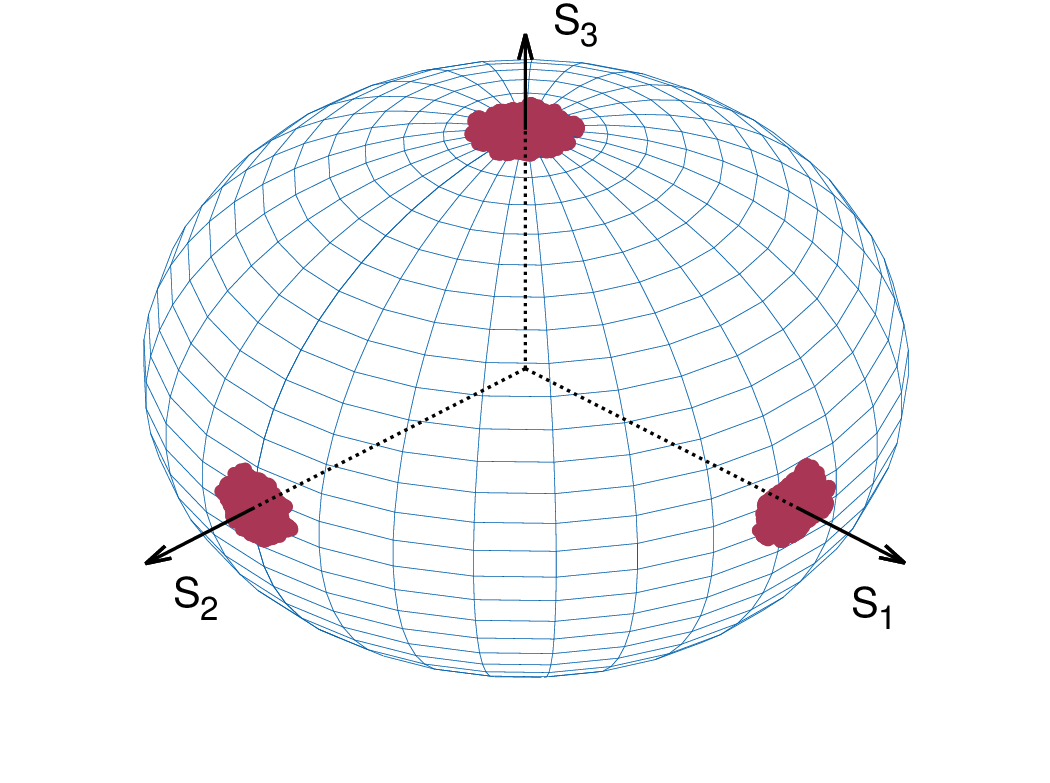}
\par\end{center}%
\end{minipage}\includegraphics[width=0.25\textwidth]{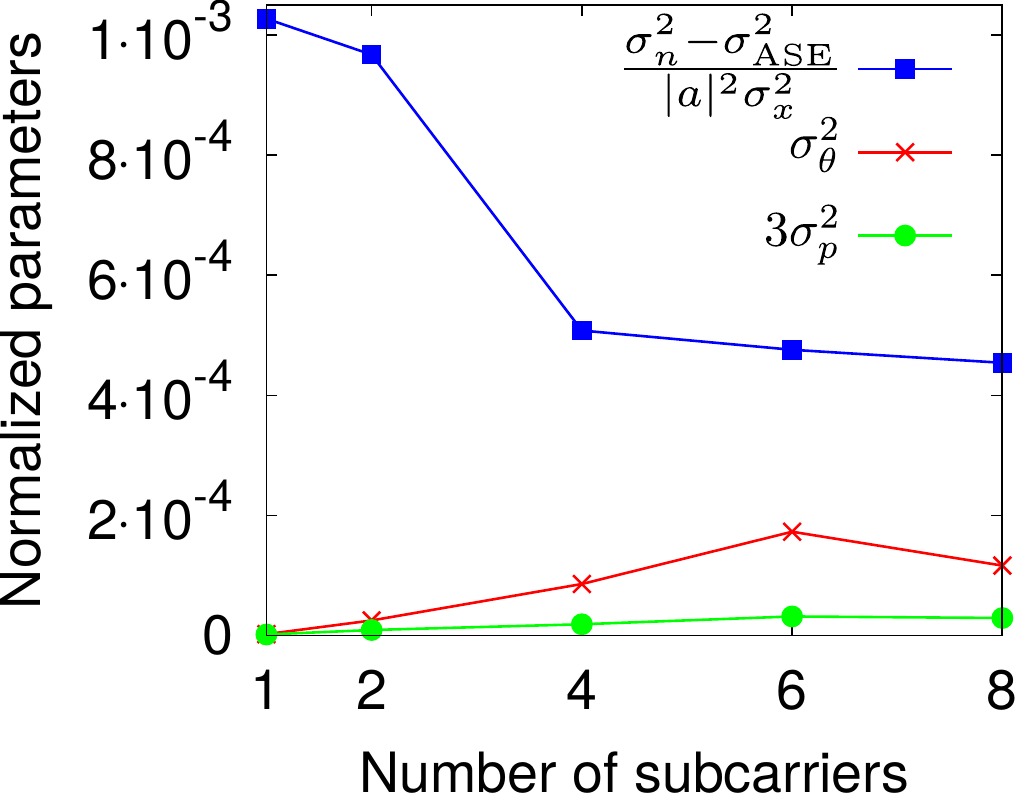}

\hspace*{0.13\textwidth}{\scriptsize{}(a)}\hspace*{0.23\textwidth}{\scriptsize{}(b)}\hspace*{0.23\textwidth}{\scriptsize{}(c)}\hspace*{0.27\textwidth}{\scriptsize{}(d)}\hspace*{\fill}

\caption{\label{fig:AIR-IDA-1000km}1000~km IDA link with 2-pol transmission
and PPN detection: (a) SE vs. launch power for different number of
subcarriers; (b) phase (and its autocorrelation in the inset) and
(c) polarization rotations induced by fiber nonlinearity at $\unit[-7]{dBm}$
and $N=4;$ (d) estimated parameters of the PPN auxiliary channel
vs. $N$ at $\unit[-7]{dBm}$.}
\end{figure*}

First, we consider a 1000~km IDA link with a 2-pol transmission (the
1-pol case was shown in \cite{secondini:ecoc17}). Fig.~\ref{fig:AIR-IDA-1000km}(a)
shows the SE as a function of the launch power (per each polarization
of each WDM channel) for the PPN detector and different numbers of
subcarriers $N$. As a comparison, we also report the SE obtained
with the AWGN detector. Note that, for CSCG input symbols, changing
$N$ does not alter the input distribution. As a consequence, also
the amount of generated NLI, the AIR with the AWGN detector, and the
(unknown) information rate (\ref{eq:MI}) remain unchanged. However,
when considering the PPN detector, the mismatch between channel and
detector depends on $N$. In fact, ISI and ICI, unaccounted for by
the model, respectively decreases and increases with increasing $N$.
An optimum value of $N$ exists, which, by minimizing the overall
interference, minimizes the mismatch and maximizes the AIR and, hence,
the SE.\textcolor{red}{{} }Thanks to the large coherence time and bandwidth
of IDA links (see Fig.~\ref{fig:XPM-coherence}), a significant SE
gain (the difference between the SE values obtained by the PPN and
AWGN detectors at their respective optimal launch power) of about
0.8~bit/s/Hz/pol is achieved for $N=4$.

The gain obtained with the PPN detector is due to its capability to
accurately track the time evolution of the state of the auxiliary
channel model (\ref{eq:PPN-auxchannel}), represented by the phase
$\theta_{k}$ and the rotation matrix $\mathsf{J}_{k}$. An example
of such evolution, extracted as a byproduct of the AIR computation
algorithm,\footnote{At each sampling time, the algorithm generates a particle representation
of the state distribution. The actual estimate is then computed as
the expectation over the particles.} is represented in Fig.~\ref{fig:AIR-IDA-1000km}(b) and (c) for
one of the central subcarriers of a 4-subcarrier system and an optimum
launch power of $\unit[-7]{dBm}$. In particular, Fig.~\ref{fig:AIR-IDA-1000km}(b)
reports the evolution of the phase $\theta_{k}$ over 20000 symbol
times (together with a zoomed version and its estimated autocorrelation
function), while the red points on the Poincaré sphere in Fig.~\ref{fig:AIR-IDA-1000km}(c)
represent the rotations of the Cartesian basis vectors $S_{1}$, $S_{2}$,
and $S_{3}$ in the Stokes space induced by the estimated $\mathsf{J}_{k}$
over the same time interval. As anticipated in Sections~\ref{subsec:Phase-noise-model}
and \ref{subsec:Phase-and-polarization}, both $\theta_{k}$ and $\mathsf{J}_{k}$
do not evolve as true random walks, as postulated in (\ref{eq:theta_transition_probability})
and (\ref{eq:rotation_matrix}), but rather show limited fluctuations
around their mean value, being therefore more similar to AR processes.
Despite such an apparent mismatch with respect to the true channel,
the adopted PPN model provides a relevant gain in terms of achievable
SE, thanks to the significant temporal correlation of the phase values
(and of the polarization rotations, whose correlation is not shown
in the figure). The possibility to get higher gains by adopting a
more refined channel model\textemdash which describes more accurately
the PPN evolution observed in Fig.~\ref{fig:AIR-IDA-1000km}(b)\textendash (c)
and the coherence properties shown in Fig.~\ref{fig:XPM-coherence}\textemdash is
left as a future work.

For the same configuration as in Fig.~\ref{fig:AIR-IDA-1000km}(a)\textendash (c)
and a fixed launch power of $\unit[-7]{dBm}$, Fig.~\ref{fig:AIR-IDA-1000km}(d)
shows how the characteristics of the channel seen by one of the central
subcarriers\textemdash the estimated parameters $\sigma_{n}^{2}$
(reduced by the contribution $\sigma_{\mathrm{ASE}}^{2}$ of the linearly
accumulated amplifier noise and normalized to the signal power $|a|^{2}\sigma_{x}^{2}$),
$\sigma_{\theta}^{2}$, and $\sigma_{p}^{2}$ (tripled to account
for the approximately triple impact that the three polarization rotation
terms $\alpha_{1,k}$, $\alpha_{2,k}$, and $\alpha_{3,k}$ have on
the signal compared to the PN term $\theta_{k}$) of the PPN model
(\ref{eq:PPN-auxchannel})\textemdash depend on $N$. By increasing
$N$, the frequency coherence of PPN over each subcarrier increases
(as the bandwidth becomes narrower), so that a larger portion of NLI
can be actually represented as pure (frequency-independent) PPN. This
explains the decrease of $\sigma_{n}^{2}$ and, in part, the increase
of $\sigma_{\theta}^{2}$ and $\sigma_{p}^{2}$. The latter is also
due to the decrease of the symbol rate of each subcarrier, which reduces
the temporal coherence of PPN over a symbol time. The best tradeoff
is obtained for $N=4$ subcarriers, corresponding to the highest SE
in Fig.~\ref{fig:AIR-IDA-1000km}(a).

\begin{figure}
\begin{centering}
\includegraphics[width=0.5\columnwidth]{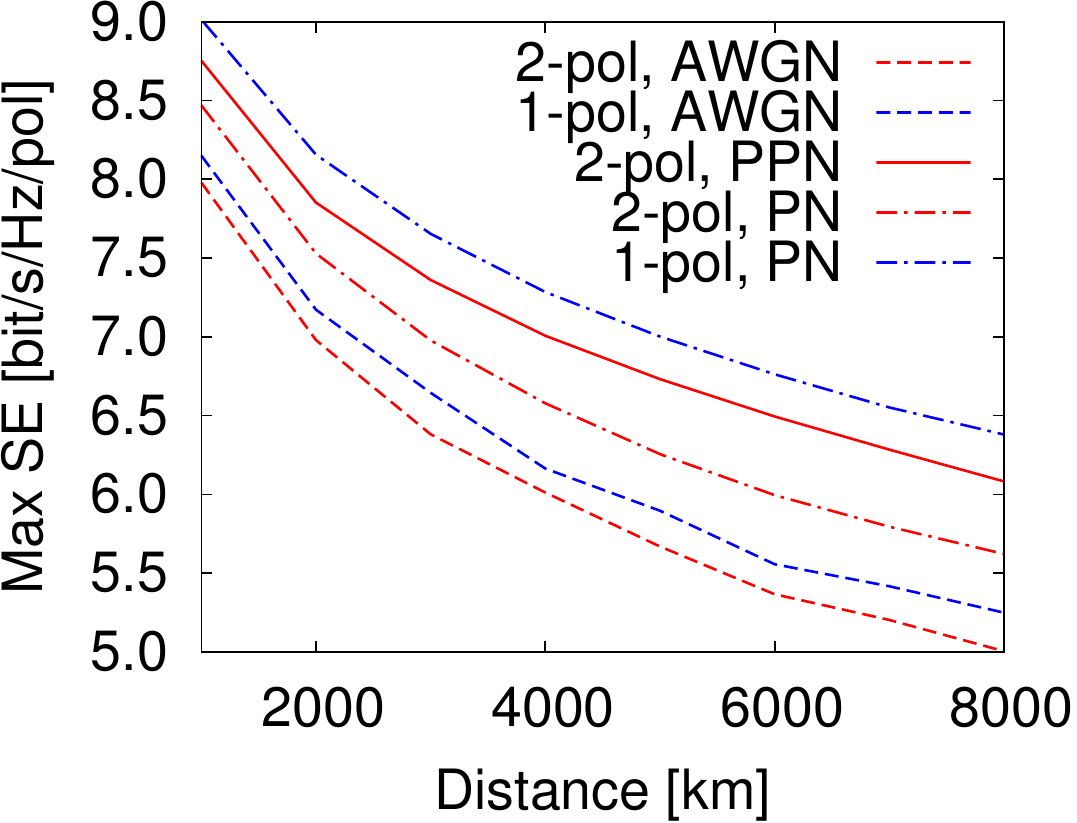}\includegraphics[width=0.5\columnwidth]{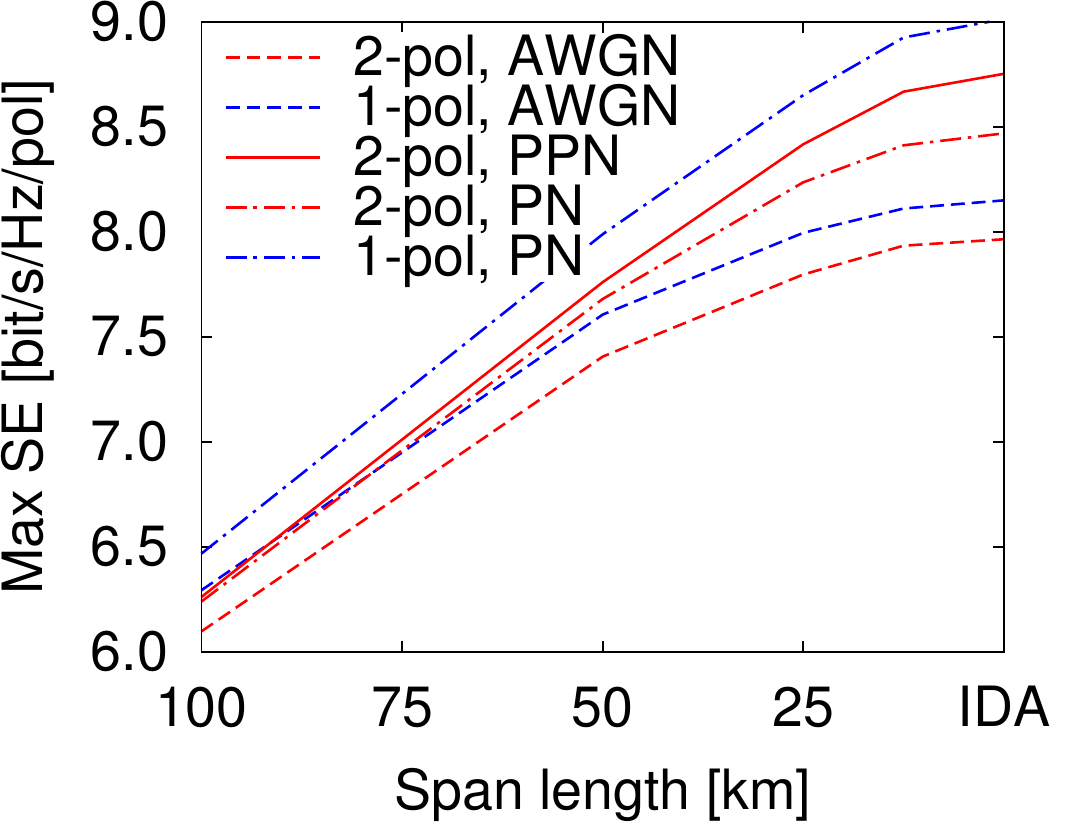}
\par\end{centering}
\hspace*{0.25\columnwidth}{\scriptsize{}(a)}\hspace*{0.48\columnwidth}{\scriptsize{}(b)}\hspace*{\fill}

\caption{\label{fig:maximumSE_IDAlink_and_1000km-link}Dependence of the achievable
SE on: (a) distance, for an IDA link; (b) amplifier spacing for a
1000~km LA link.}
\end{figure}

The maximum achievable SE\textemdash i.e., the SE obtained by optimizing
the launch power and the number of subcarriers\textemdash depends
on the link and system configuration and on the detection strategy.
For instance, Fig.~\ref{fig:maximumSE_IDAlink_and_1000km-link} shows
the maximum SE achievable by different detectors in DU links as a
function of (a) the link length (for an IDA link) and (b) the amplifier
spacing (for a 1000~km LA link). For the sake of comparison, both
1-pol and 2-pol transmissions are considered. For IDA links, relevant
gains are obtained both in the 1-pol and 2-pol scenarios by the PN
and PPN detectors, respectively. At 4000~km, the SE gain is about
1.1~bit/s/Hz in the 1-pol case, and 1~bit/s/Hz/pol in the 2-pol
case with PPN detection. In the latter case, the gain reduces to about
0.6~bit/s/Hz/pol if the two polarizations are independently processed
by PN detection, neglecting polarization effects. The gains slightly
increase for longer distances, and appear even more relevant if measured
in terms of reach: for a target SE of 7~bit/s/Hz/pol, the reach is
doubled in the 2-pol case (with PPN detection), and more than doubled
in the 1-pol case. On the other hand, lower gains are obtained in
LA links, due to their lower time and frequency coherence (see, for
instance, Fig.~\ref{fig:XPM-coherence}). For the 1000~km link considered
in Fig.~\ref{fig:maximumSE_IDAlink_and_1000km-link}(b), the SE gain
in the 2-pol case gradually decreases from 0.8 to 0.2~bit/s/Hz/pol
when increasing the amplifier spacing from zero (actually corresponding
to an IDA configuration) to 100~km. A similar behavior is observed
in the 1-pol case.

\begin{figure}
\begin{centering}
\includegraphics[width=0.5\columnwidth]{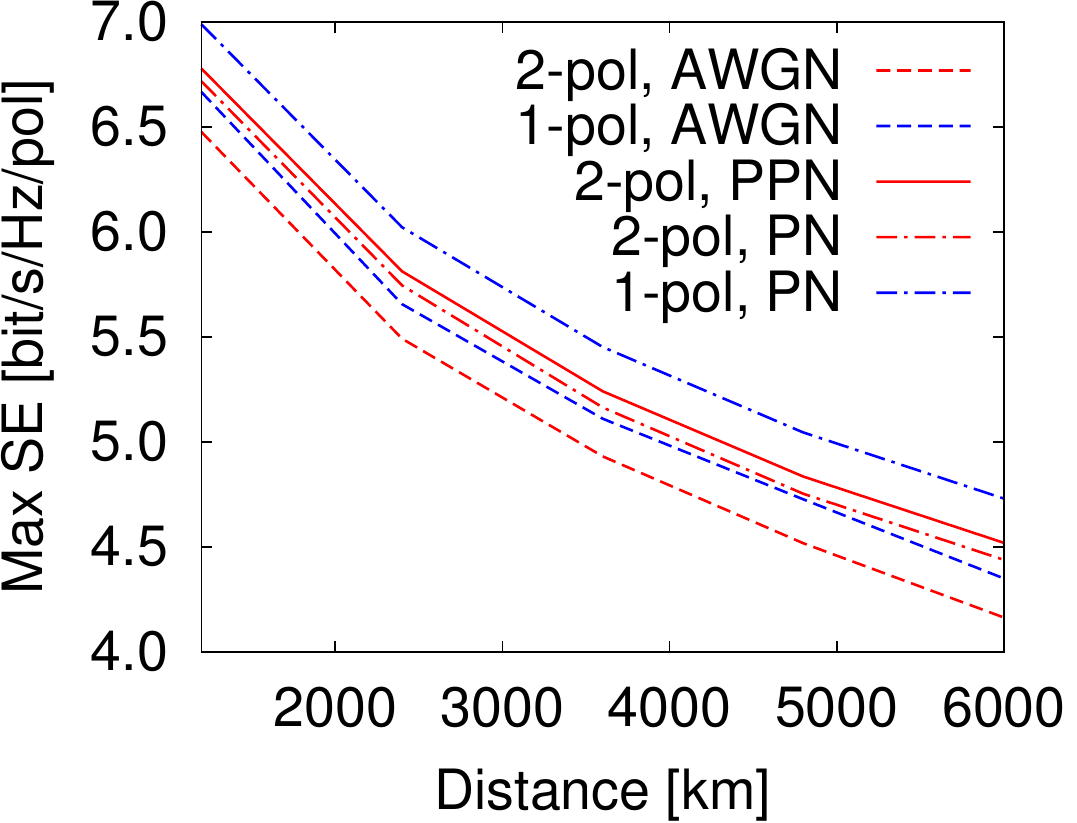}\includegraphics[width=0.5\columnwidth]{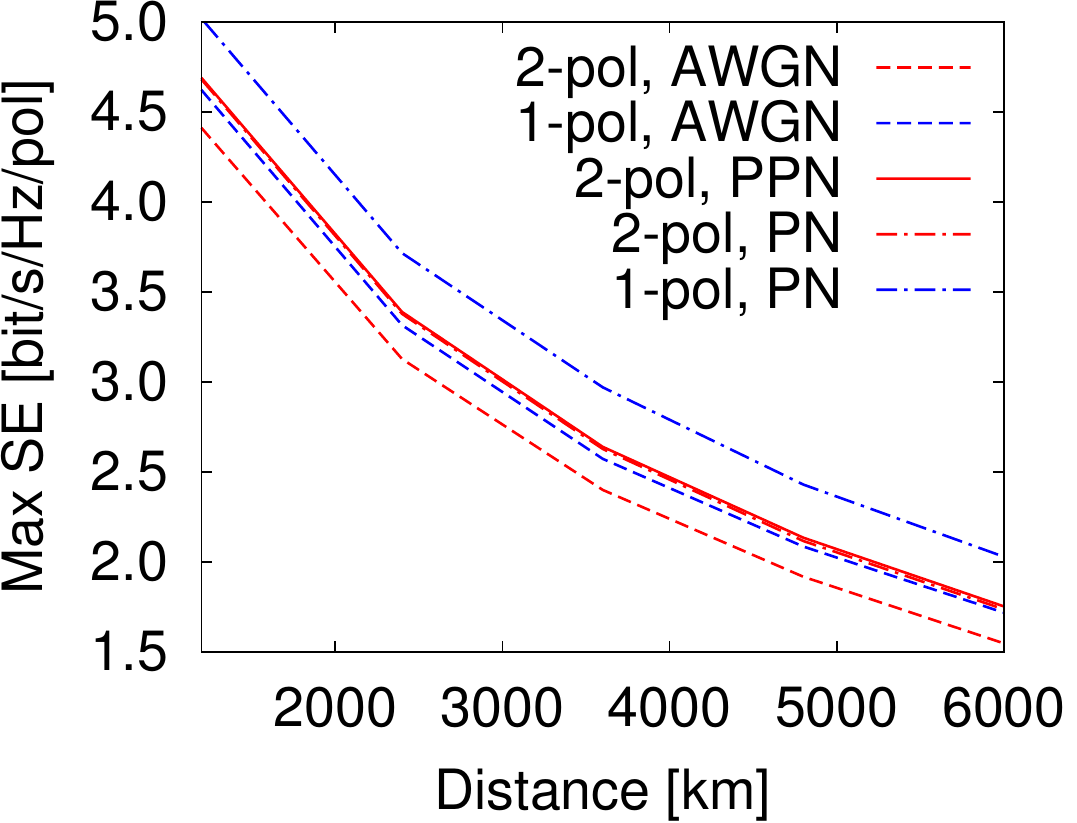}
\par\end{centering}
\hspace*{0.25\columnwidth}{\scriptsize{}(a)}\hspace*{0.48\columnwidth}{\scriptsize{}(b)}\hspace*{\fill}

\caption{\label{fig:SEgain_DUandDMlinks}Dependence of the achievable SE on
distance for an $N_{s}\times\unit[60]{km}$ (a) LA-DU and (b) LA-DM
link.}
\end{figure}

Fig.~\ref{fig:SEgain_DUandDMlinks} provides the same comparison
as in Fig.~\ref{fig:maximumSE_IDAlink_and_1000km-link}(a) but for
an $N_{s}\times\unit[60]{km}$ (a) LA-DU and (b) LA dispersion-managed
(DM) link. In the DM link, dispersion is fully compensated after each
span by $\unit[10.2]{km}$ of dispersion-compensating fiber (DCF)
with parameters $\alpha=\unit[0.57]{dB/km}$, $\beta_{2}=\unit[127.5]{ps^{2}/km}$,
and $\gamma=\unit[6.5]{W^{-1}\cdot km^{-1}}$. An additional optical
amplifier with $\eta=1.6$ is inserted at the input of each DCF to
set the launch power in the DCF always at 4~dB below that in the
transmission fiber. For the DU link, at 4800~km, an SE gain of about
0.3~bit/s/Hz/pol is obtained in both the 1-pol and 2-pol cases, which
can be alternatively exploited to increase the reach by 25\% (up to
6000~km) keeping the SE fixed. The gain reduces to 0.2~bit/s/Hz/pol
(+20\% reach) in the 2-pol case if independent PN detection on each
polarization is employed. On the other hand, in the DM link, the gain
provided by the PPN strategy is reduced (about 0.2~bit/s/Hz/pol at
4800~km) due to the lower temporal coherence. This is in contrast
with the predictions made in \cite{dar_JLT2017_nonlinear} in terms
of potential peak SNR gain for single-span links, which should behave
similarly to DM links. As mentioned in \cite{dar_JLT2017_nonlinear},
those ``potential'' SNR gains are based on ideal PPN removal and
do not account for the temporal correlation of PPN, which is quite
short in DM links. As a result, they cannot be fully translated into
\emph{achievable} SE gains.

The optimum number of subcarriers considered in the previous cases
depends on the link configuration: it increases with distance from
4 to 8 in Fig.~\ref{fig:maximumSE_IDAlink_and_1000km-link}(a) and
from 2 to 6 in Fig.~\ref{fig:SEgain_DUandDMlinks}(a), while it is
$N=1$ at any distance in Fig.~\ref{fig:SEgain_DUandDMlinks}(b).

\section{Discussion and conclusions}

We have discussed the importance of channel models for nonlinearity
mitigation in WDM fiber-optic systems, and the need to drop the AWGN
assumption to devise improved detection strategies. A step toward
this direction is done by resorting to the FRLP model, which suggests
considering interchannel NLI as linear doubly dispersive (in time
and in frequency) fading. This analogy helps to understand the characteristics
of the channel in terms of its coherence time and bandwidth, and to
devise improved transmission schemes borrowed from wireless communications.
On this basis, we have derived some simple discrete-time auxiliary
channel models and employed SCM, optimizing the number of subcarriers,
to maximize the AIR and found some improved SE lower bounds. For IDA
links, thanks to the high time\textendash frequency coherence of interchannel
NLI, the detection strategies based on the considered channel models
provide about 1~bit/s/Hz/pol SE gain at 4000~km, and double the
reach for a fixed SE of 7~bit/s/Hz/pol compared to a detection based
on the AWGN model. The gain increases with distance. In LA links,
the coherence time and bandwidth of interchannel NLI are reduced,
so that lower gains are obtained. The gain increases with the number
of spans, decreases with the span length, and is further reduced by
in-line dispersion compensation. Similar behaviors are observed in
1-pol and 2-pol systems, provided that, in the latter, the two polarizations
are jointly detected to account for polarization effects. All the
gains were measured for a 5-channel WDM system to keep the computational
complexity low, but are expected to increase when more channels are
present \cite{Sec:PTL14}.

Another interesting result of this study is the observed synergy of
SCM and PPN compensation. When detection is optimized for the AWGN
channel, the benefits provided by SCM in terms of nonlinearity mitigation
are reduced for high-order QAM modulations, and vanish as the modulation
alphabet approaches a Gaussian distribution \cite{poggiolini:JLT16}.
On the other hand, when employing PN or PPN detection, SCM allows
to obtain the best trade-off between coherence time and bandwidth
by optimizing the number of subcarriers, hence fully revealing the
NLI mitigation capabilities of these detection strategies and maximizing
the achievable SE and reach.

This study also suggests that further optimizations and gains are
possible, which are left as a future work. Potential improvements
include, for instance, accounting for phase and polarization correlation
among subcarriers or considering higher-order AR processes to model
the evolution of the channel state. This can be done either analytically,
based on the FRLP model, or from a statistical analysis of the HMM
parameters based on simulated/experimental data. Finally, unlike
conventional fading, the channel response derived from the FRLP model
does not depend on environmental conditions, out of the user control
(e.g., multipath and shadowing effects), but rather on the actual
signals transmitted by the other users (channels, according to WDM
terminology). Even excluding unpractical joint modulation or detection,
this peculiarity gives the possibility of controlling the channel
response by adopting a proper combination of coding and modulation
on each channel. In this context, an interesting but still unexplored
possibility is that of using constellation shaping to increase channel
coherence, rather than to decrease NLI. This approach might have a
better synergy with the ones discussed in this work and provide additional
SE gains.\bibliographystyle{ieeetr}
\bibliography{refs}

\end{document}